\documentclass[a4paper,fleqn,usenatbib]{mnras}

\usepackage[T1]{fontenc}
\usepackage{ae,aecompl}

\usepackage{graphicx}	
\usepackage{epstopdf}
\usepackage{savesym}
\usepackage{amsmath}	
\usepackage{amssymb}	
\savesymbol{iint}
\usepackage{txfonts}
\restoresymbol{TXF}{iint}

\title[GAMA: AGN in Pairs of Galaxies]{Galaxy And Mass Assembly (GAMA): Active Galactic Nuclei in Pairs of Galaxies}

\author[Y.A. Gordon et al.]{Yjan A. Gordon$^{1,2}$\thanks{E-mail: y.gordon@hull.ac.uk (YAG)},
Matt S. Owers$^{3,4}$,
Kevin A. Pimbblet$^{1,2,5,6}$,
Scott M. Croom$^{7}$,
\newauthor Mehmet Alpaslan$^{8}$,
Ivan K. Baldry$^{9}$
Sarah Brough$^{4}$,
Michael J.I. Brown$^{5,6}$
\newauthor Michelle E. Cluver$^{10}$,
Christopher J. Conselice$^{11}$,
Luke J.M. Davies$^{12}$,
\newauthor Benne W. Holwerda$^{13}$,
Andrew M. Hopkins$^{4}$,
Madusha L.P. Gunawardhana$^{14}$
\newauthor Jonathan Loveday$^{15}$, 
Edward N. Taylor$^{16}$
and Lingyu Wang$^{17, 18}$
\\
$^{1}$E.A. Milne Centre for Astrophysics, University of Hull, Cottingham Road, Kingston-upon-Hull, HU6 7RX, UK\\
$^{2}$School of Mathematics and Physical Sciences, University of Hull, Cottingham Road, Kingston-upon-Hull, HU6 7RX, UK\\
$^{3}$Department of Physics and Astronomy, Macquarie University, NSW 2109, Australia\\
$^{4}$Australian Astronomical Observatory (AAO), PO Box 915, North Ryde, NSW 1670, Australia\\
$^{5}$Monash Centre for Astrophysics (MoCA), Monash University, Clayton, Victoria 3800, Australia\\
$^{6}$School of Physics and Astronomy, Monash University, Clayton, Victoria 3800, Australia\\
$^{7}$Sydney Institute for Astronomy (SIfA), School of Physics, University of Sydney, NSW 2006, Australia\\
$^{8}$NASA Ames Research Centre, N232, Moffett Field, Mountain View, CA 94035, USA\\
$^{9}$Astrophysics Research Institute, Liverpool John Moores University, IC2, Liverpool Science Park, 146 Brownlow Hill, Liverpool, L3 5RF, UK\\
$^{10}$Department of Physics and Astronomy, University of the Western Cape, Robert Sobukwe Road, Bellville, 7535, South Africa\\
$^{11}$School of Physics and Astronomy, University of Nottingham, Nottingham, NG7 2RD, UK\\
$^{12}$ICRAR, The University of Western Australia, 35 Stirling Highway, Crawley, WA 6009, Australia\\
$^{13}$University of Leiden, Sterrenwacht Leiden, Niels Bohrweg 2, NL-2333 CA Leiden, The Netherlands\\
$^{14}$Instituto de Astrof\'isica and Centro de Astroingenier\'ia, Facultad de F\'isica, Ponticia Universidad Cat\'olica de Chile, Vicu\~na Mackenna 4860,\\ \hspace{2mm} 7820436 Macul, Santiago, Chile\\
$^{15}$Astronomy Centre, University of Sussex, Falmer, Brighton, BN1 9QH, UK\\
$^{16}$Centre for Astrophysics and Supercomputing, Swinburne University of Technology, Hawthorn 3122, Australia\\
$^{17}$SRON Netherlands Institute for Space Research, Landleven 12, 9747 AD, Groningen, The Netherlands\\
$^{18}$Kapteyn Astronomical Institute, University of Groningen, Postbus 800, 9700 AV Groningen, the Netherlands
}

\date{Accepted XXX. Received YYY; in original form ZZZ}

\pubyear{2016}

\begin{document}
\label{firstpage}
\pagerange{\pageref{firstpage}--\pageref{lastpage}}
\maketitle

\begin{abstract}
There exist conflicting observations on whether or not the environment of broad and narrow line AGN differ and this consequently questions the validity of the AGN unification model. The high spectroscopic completeness of the GAMA survey makes it ideal for a comprehensive analysis of the close environment of galaxies. To exploit this, and conduct a comparative analysis of the environment of broad and narrow line AGN within GAMA, we use a double-Gaussian emission line fitting method to model the more complex line profiles associated with broad line AGN. We select 209 type 1 (i.e., unobscured), 464 type 1.5-1.9 (partially obscured), and 281 type 2 (obscured) AGN within the GAMA II database. Comparing the fractions of these with neighbouring galaxies out to a pair separation of $350\,\text{kpc }h^{-1}$ and $\Delta z < 0.012$ shows no difference between AGN of different type, except at separations less than $20\,\text{kpc }h^{-1}$ where our observations suggest an excess of type 2 AGN in close pairs. We analyse the properties of the galaxies neighbouring our AGN and find no significant differences in colour or the star formation activity of these galaxies. Further to this we find that $\Sigma_5$ is also consistent between broad and narrow line AGN. We conclude that the observations presented here are consistent with AGN unification.
\end{abstract}




\section{Introduction}
\label{s1}
Type 1 and Type 2 (T1 and T2 respectively hereafter) active galactic nuclei (AGN) are defined by the presence or absence of broad emission lines in their spectra, respectively. When \citet{b3} discovered the presence of hidden broad emission lines in the polarized spectra of T2 Seyfert galaxies it was suggested that a dusty torus around the active nucleus was responsible for scattering the light, resulting in the so called ``unified model of AGN". In the most simple interpretation of AGN unification, both T1 and T2 AGN are expected to be the same type of object and only the orientation of a circumnuclear dusty torus differs relative to the observer \citep{b4}. If the unified model is a complete description then all T2 AGN should contain hidden broad lines. Problematically for AGN unification, these have only been discovered in approximately 50\% of T2 AGN \citep{b1}. This suggests that at least some T2 AGN are fundamentally different to T1 AGN. However, the lack of hidden broad lines in all T2 AGN might be solely explained by variable homogeneity and covering or obscuration factor of the dusty torus around the central engine \citep{b13}.

If the differences in observed properties of T1 and T2 AGN are simply due to the orientation of the torus with respect to viewing angle, as predicted in the AGN unification scheme, then there should be no significant difference between the external environment of these two types of AGN. However, there have been observations that demonstrate that T1 and T2 Seyfert galaxies (Sy1 and Sy2 respectively hereafter) are not found in identical environments. \citet{b39}, \citet{b40} and \citet{b35} have all found that Sy2 galaxies are significantly more likely than Sy1 galaxies to be in a galaxy pair with a projected separation of less than 100 kpc $h^{-1}$. Furthermore, \citet{b41} found that Sy2s reside in similar environments to galaxies in the IRAS bright galaxy sample \citep{b42, b43} whereas Sy1s do not. \citet{b41} and \citet{b40} suggest that this may imply an evolutionary path as a result of galaxy interactions from star-forming, to Sy2 through to Sy1 post interaction, and thus implying that AGN unification is an inaccurate model.

As well as the simple likelihood of an AGN to be of a particular type dependent upon the presence or absence of a nearby galaxy, \citet{b2} have conducted a thorough census of the properties of AGN neighbours. By using the Sloan Digital Sky Survey (SDSS; \citealp{b6}) seventh data release (DR7; \citealp{b8}) they found that in AGN-AGN pairs the neighbour of a T2 AGN was significantly more likely to also be a T2 AGN at projected separations below 200 kpc, further strengthening the validity of prior observations (e.g, \citealp{b39, b40, b44, b35}). Further to this \citet{b2} found that the colour of the neighbouring galaxies differed with AGN type; the neighbours of T1s being redder than the neighbours of T2s. This implies that either the star-formation rate, metallicity or stellar population age of the neighbouring galaxy may affect the type of AGN triggered by the interaction.

As mergers are believed to be a primary source of fuel for AGN \citep{b5}, studying interacting galaxies where one galaxy hosts an AGN provides an opportunity to investigate how nuclear activity evolves from its earliest stages. As interacting galaxies appear in the sky as close pairs, a major drawback facing some previous studies using projects such as the SDSS (e.g., \citealp{b2}) is the inability to detect the closest pairs of galaxies as a result of spectroscopic fibre collisions \citep{b7}. In these situations, including the work by \citet{b2}, photometric data is often used to support the spectroscopic observations. This can result in less accurate $z$-space separations of the pairs being measured and a lack of emission line data to classify potential AGN. SDSS DR7 \citep{b6,b8} contains about 1,050,000 galaxy spectra accounting for 94\% of potential targets with the remainder being lost to fibre collisions \citep{b9}. Indeed, in dense regions of the sky, fibre collisions severely hamper the completeness of the SDSS spectroscopic catalogue \citep{b24}.

Surveys such as the Galaxy And Mass Assembly (GAMA) survey \citep{b10, b46} circumvent the fibre collision problem by observing each field of view multiple times, moving the fibres between observations \citep{b11}. This results in a far more complete sample of close pairs with which to study galaxy interactions than is possible in spectroscopic surveys that do not use this technique. Indeed, the completeness of GAMA has already been used to better constrain the effect of close environment and galaxy interactions on galaxy evolution \citep{b67, b45, b66, b15, b68}. Given the questions asked of AGN in unification from observations over the last few decades, we aim to further scrutinise this widely accepted model by taking advantage of the high spectroscopic completeness of GAMA in order to thoroughly test the environment of broad and narrow line AGN. Furthermore, we investigate whether or not AGN with a single broad Balmer emission are indeed the same as AGN with multiple broad Balmer emission lines by comparing how these populations behave under a range of environmental probes.

In Section \ref{s2} of this paper we describe the spectral emission line fitting method used to produce the data set from which we select our AGN. Section \ref{s3} outlines our method of selecting AGN using the resultant emission line catalogue and details how we select galaxy pairs. In Section \ref{s4} we analyse the environment of our AGN and discuss our observations which we compare to other observations in Section \ref{S_comp}. Our conclusions are stated in Section \ref{s5}. Throughout this paper we use a standard flat $\Lambda$CDM cosmology: $h=0.7,\ H_{0} = 100h\ \rm{km}\ \rm{s}^{-1} \rm{Mpc}^{-1},\ \Omega_{M} = 0.3,\ \Omega_{\Lambda} = 0.7$.

\section{Data}
\label{s2}
The GAMA survey was undertaken between 2008 and 2014 \citep{b10, b46} at the Australian Astronomical Observatory (AAO) using the 3.9m Anglo-Australian Telescope. GAMA obtained spectra for $> 250000$ galaxies with $r<19.8\,\text{mag}$ using the 2dF/AAOmega spectrograph \citep{b30}. For each target the 2dF/AAOmega spectrograph obtains a blue spectrum covering the range 3750-5850\AA\ and a red spectrum covering 5650-8850\AA. These are spliced together at 5700\AA\ resulting in a total spectrum, which has a mean resolution of $R \approx 1300$, with an observed wavelength range of 3750-8850\AA\ \citep{b33}. Covering $\approx 260$ square degrees across 5 regions (G02, G09, G12, G15 and G23) GAMA is $>98\%$ complete in the 3 equatorial regions G09, G12 and G15 \citep{b46} making observations from these regions highly valuable for observing close pairs of galaxies.

\subsection{Emission line modelling} 
To select our AGN we require accurate emission line measurements for the spectra of the galaxies within GAMA. To this end we present and then use the \textsc{SpecLineSFRv05} catalogue that provides line flux and equivalent width measurements for GAMA II spectra by modelling emission lines with either a single or double Gaussian profile. \textsc{SpecLineSFRv05} is constructed by selecting the spectra in the GAMA \textsc{SpecAllv27} dataset that have a redshift quality $nQ>1$ (indicating a redshift has been measured, for a full description of redshift quality in GAMA see \citealp{b10, b46}) and a redshift $0.002<z<1.35$. The catalogue contains 427,829 entries and includes repeated measurements for some targets. In cases where objects have several spectra, the best redshift was used for the line measurements. The catalogue excludes a small number of targets which either do not have spectra available, or were taken with the Liverpool Telescope \citep{b48}. Also excluded are additional spectra from the VVDS (VIMOS VLT Deep Survey; \citealp{b49, b50})

This version of the resultant database is the first GAMA II emission line measurements catalogue (previous versions of the GAMA emission line measurement database have been for \text{GAMA I}; \citealp{b31}) and differs from previous versions in that it also provides fits for spectra from SDSS, the 2 degree Field Galaxy Redshift Survey (2dFGRS; \citealp{b16}), the 6 degree Field Galaxy Survey (6dFGS; \citealp{b55}), the WiggleZ survey \citep{b56}, the Millennium Galaxy Catalogue (MGC; \citealp{b52}), the 2dF-SDSS LRG and QSO survey (2SLAQ; \citealp{b53, b54}) and the 2dF QSO redshift survey (2QZ; \citealp{b23}). Each spectrum is fitted across 5 regions containing 12 physically important emission lines. The continuum in each of these regions is modelled as a straight line. The fitting is done with the IDL code `mpfitfun' \citep{b14} which uses a Levenberg-Marquardt non-linear least squares minimisation to identify the best-fitting parameters for the model given the data and its associated uncertainties. The fitted spectral regions are:
\begin{itemize}
\item 3626-3779\AA\ to model the [O II] lines at 3726\AA\ and 3729\AA.
\item 4711-5157\AA\ to model H$\beta$ and the [O III] lines at 4959\AA\ and 5007\AA.
\item 6270-6394\AA\ covering the [O I] emission lines at 6300\AA\ and 6364\AA.
\item 6398-6710\AA\ modelling H$\alpha$ and the [N II] lines at 6548\AA\ and 6583\AA.
\item 6616-6831\AA\ to model [S II] lines at 6716\AA\ and 6731\AA.
\end{itemize}

The resultant catalogue is organised into 3 data tables. The first table, \textsc{SpecLineGaussFitSimple}, contains line measurements derived from single-Gaussian fits as well as the strength of the 4000\AA\ break, measured using the method of \citet{b57} using the continuum band definitions given by \citet{b58}. There is also an estimate of the S/N per pixel in the continuum measured in the 153\AA\ window from 6383 - 6536\AA. This is 12\AA\ blue-ward of the [N II](6548\AA) line and is measured on the redshifted spectrum. 

The second table, \textsc{SpecLineGaussFitComplex}, contains more complicated fits to the regions containing the H$\alpha$ and H$\beta$ lines. For H$\alpha$ and H$\beta$, a second Gaussian component is added that may be either in absorption or broad emission. Model selection scores are described below (Section \ref{s23}) and can be used to select those spectra where this extra component is justified given the improvement in the fit due to the extra components.

The third table, \textsc{SpecLineDirectSummation}, contains direct summation equivalent widths for 51 absorption and emission line species. The equivalent widths and their associated uncertainties are measured using the techniques outlined in \citet{b57}. There are no corrections made due to the effects of velocity dispersion, nor are there attempts to place Lick index measurements onto the Lick system. 

\subsection{Gaussian line fitting procedure}
\begin{figure}
\includegraphics[scale=0.74]{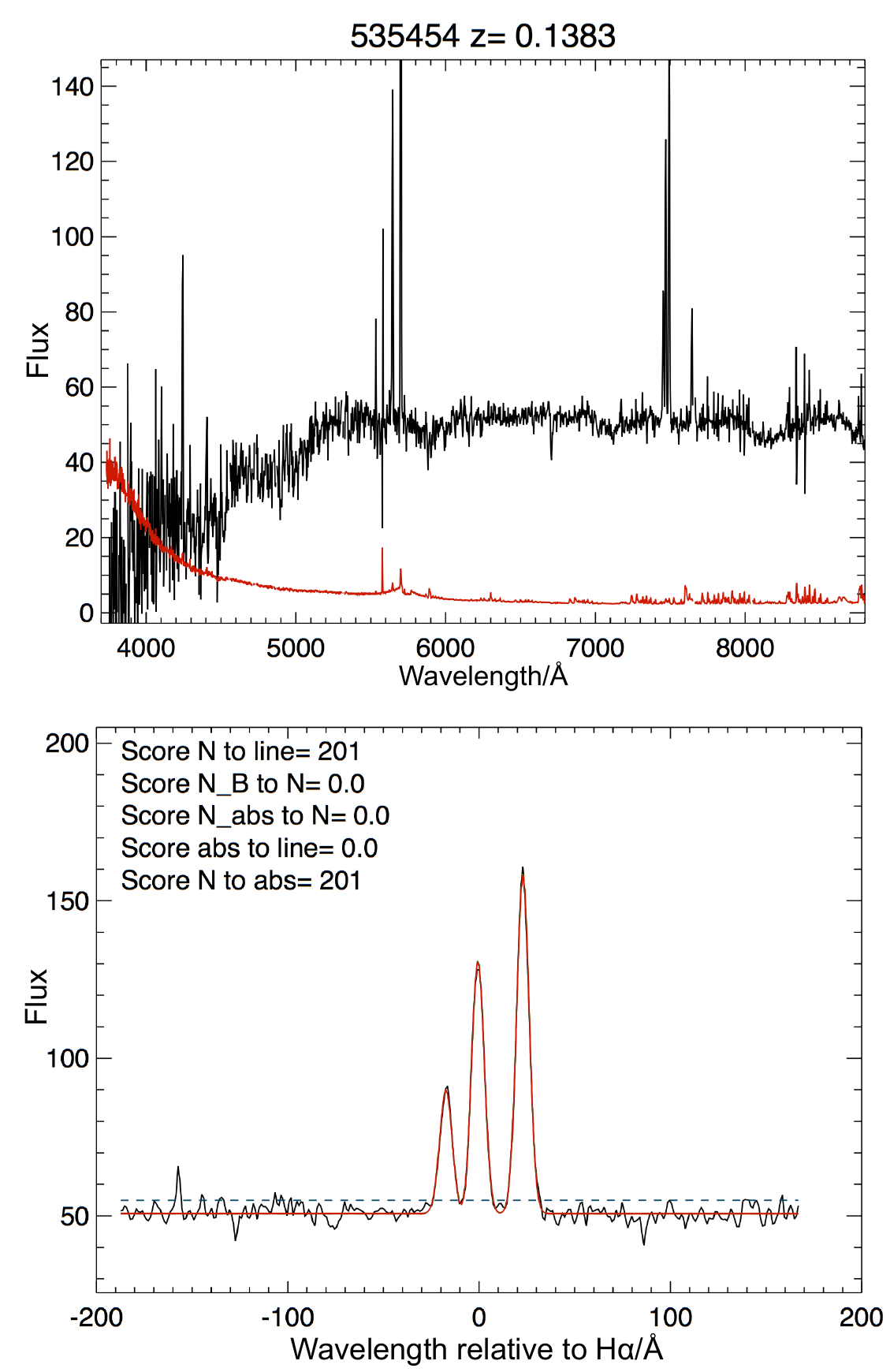}
\caption{Example of Gaussian fit to a narrow H$\alpha$ emission line. Top: AAOmega spectrum of the galaxy (GAMA CATAID = 535454). Bottom: fitting of the 6398-6710\AA\ region containing H$\alpha$ and the two [N II] lines. The scores in the top left of the bottom panel are the model selection scores (see Section \ref{s23}) for the different complexities of fitting model. `Line' is the continuum only model, `N' is the narrow emission model, `N\_B' is the narrow + broad emission model, `abs' is the absorption only model and `N\_abs' is the narrow emission + absorption model. The score of 201 involving the the narrow model and either just the continuum or  absorption `N to line' and `N to abs' of 201 indicates that this is this is the preferred fit.}
\label{HaN}
\end{figure}

\begin{figure}
\includegraphics[scale=0.75]{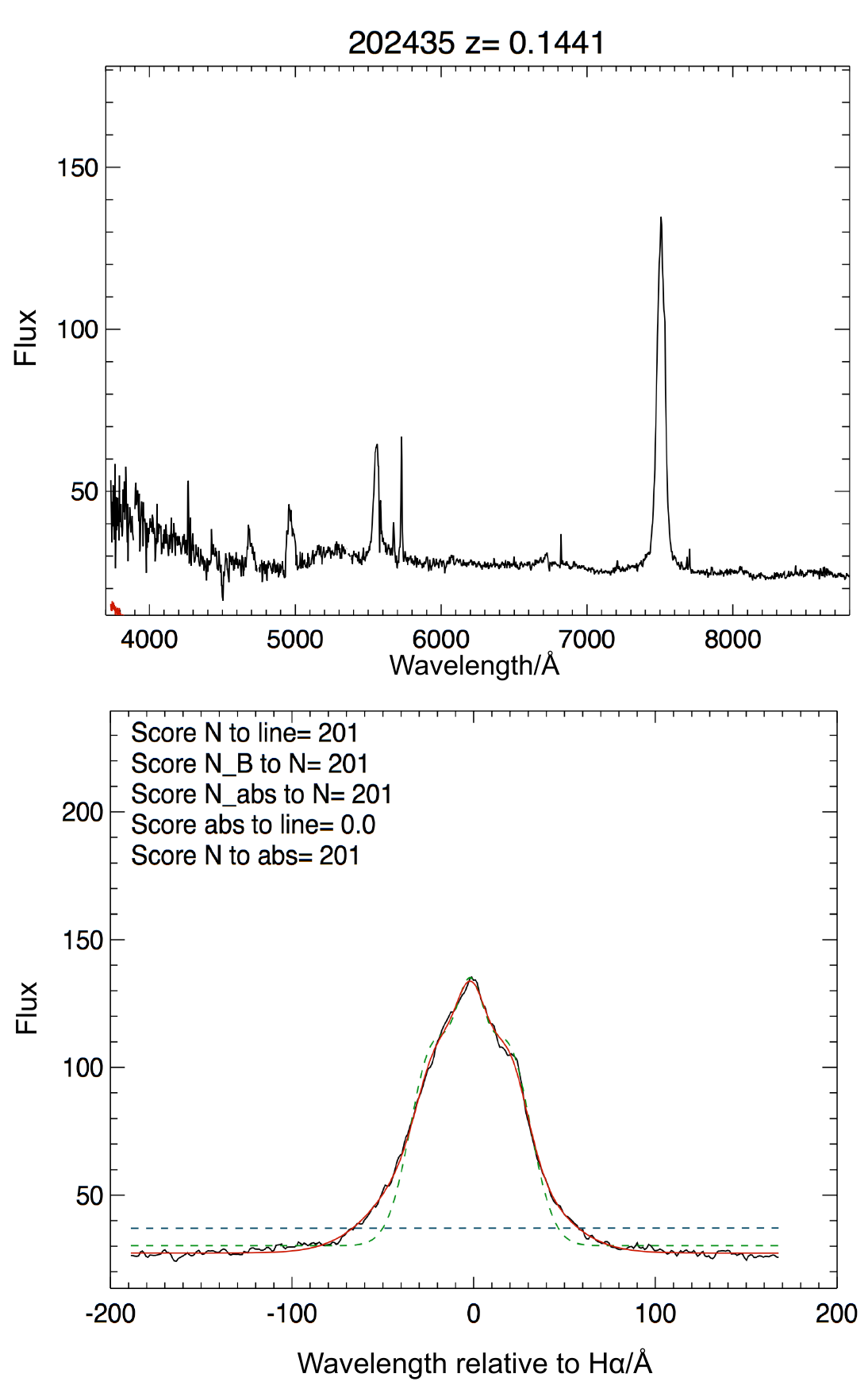}
\caption{Example of Gaussian fit to a broad H$\alpha$ emission line. Top: AAOmega spectrum of the galaxy (GAMA CATAID = 202435). Bottom: fitting of the 6398-6710\AA\ region containing H$\alpha$ and the two [N II] lines. The score of 201 involving the the most complex model `N\_B to N' of 201 indicates that this is this is the preferred fit (see Section \ref{s23}).}
\label{HaB}
\end{figure}

The fitting begins with a simple straight line fit to the spectral region of interest (listed above) and increases in complexity depending on the line species. Regions containing lines that are expected to only occur as narrow emission lines (e.g., [O II], [S II] and [O I]) have only one level of complexity above that of a straight line fit (the inclusion of the Gaussian lines for the narrow emission). For the regions containing H$\alpha$ and H$\beta$, there are six different manifestations of absorption and emission. For example, the \text{H$\alpha$ + [N II](6548/6583\AA)} region can contain the following combinations in increasing complexity (likewise for H$\beta$ + [O III](4959/5007\AA):
\begin{itemize}
\item No emission or absorption, just continuum.
\item H$\alpha$ in absorption and no [N II] emission.
\item\relax [N II](6548\AA) + H$\alpha$ + [N II](6583\AA) all with narrow emission.
\item\relax [N II](6548\AA) + [N II](6583\AA) in emission + H$\alpha$ in absorption.
\item\relax [N II](6548\AA) + [N II](6583\AA) in emission + H$\alpha$ in emission and absorption.
\item\relax [N II](6548\AA) + [N II](6583\AA) in emission + H$\alpha$ in narrow plus broad emission.
\end{itemize}
Each of the above fits are performed on the data and a model selection score is given for the more complex model compared with the simpler one (see below). For models with the same number of fitted parameters, the model with the lowest $\chi^{2}$ value is chosen. Examples of this line fitting for narrow and broad emission lines are shown in Figures \ref{HaN} and \ref{HaB}.


In order to fit the model to the line region the following limits are set:
\begin{itemize}
\item Line position is limited to be within 200 km s$^{-1}$ of the expected position given the redshift of the galaxy (the redshift has the heliocentric correction removed and SDSS spectra are converted from vacuum to air wavelengths).
\item For the narrow emission line components, the width of the Gaussian, $\sigma$, is constrained to be in the range $0.75\sigma_{\text{inst}} < \sigma <\sqrt{500^2+\sigma_{ \text{inst}}^2}$ where $\sigma_{\text{inst}}$ is the instrumental resolution of the spectrum in $\text{km s}^{-1}$. This constrains the width of the narrow-line components to be less than $500 \text{km s}^{-1}$.
\item The boundaries on the amplitude for the line are estimated from the range in data near the expected position of the line. A small negative value is allowed for emission-only lines (e.g., [N II], [O II] etc.) in order to assess line detection limits.
\item For broad emission lines, $\sigma$ is constrained to be in the range $\sqrt{500^2 + \sigma_{\text{inst}}^2} < \sigma < \sqrt{5000^2 + \sigma_{\text{inst}}^2}$. A larger parameter space is allowed for the position of the broad component (400 kms$^{-1}$). The initial guess for intrinsic dispersion is 1000 km s$^{-1}$.
\item For all doublet lines, the position and velocity dispersion of the weaker line is tied to that of the stronger line. Given the [O II] doublet is rarely resolved, the amplitudes for [O IIB\&R] are tied to the ratio $\text{[O IIB]}=0.35\times \text{[O IIR]}$.
\end{itemize}

This method provides equivalent width (EW) and flux (F) for the fitted lines and are derived by:
\begin{equation}
EW = \frac{F}{C} = \frac{\sqrt{2\pi}A\sigma}{C}
\end{equation}
Where $A$ is the amplitude of the Gaussian, $\sigma$ is the line dispersion (including instrument dispersion) and $C$ is the continuum at the position of the emission line given by the equation of the linear fit to the continuum. The uncertainties on measurements are propagated in quadrature from the errors of their dependences. That is to say for $EW$, $\Delta EW = |EW|\times \sqrt{(\Delta A/A)^2 + (\Delta \sigma/\sigma)^2 + (\Delta C/C)^2}$. All the equivalent widths are corrected by $(1+z)$ and are thus rest-frame measurements.

\subsection{Model selection}
\label{s23}
Given the increasing complexity of the models, it is important to ensure that the data are not being over-fit because of the extra freedom allowed by the additional model parameters. There are many model selection methods, each with its own advantages and drawbacks. In an attempt to overcome the drawbacks of different methods, three model selection methods have been used to give a single model selection score. Two of the model selection methods have their roots in Bayesian statistics and are estimators of the `Bayes Factor'. Since a full Bayesian approach would be rather time-consuming for ~400,000 spectra, two analytic approximations are used to estimate the Bayes Factor. The first is the change in the Bayesian information criterion ($\Delta \text{BIC}$), where the BIC for each model is given by:
\begin{equation}
\text{BIC} = \chi^{2} +d(\ln{N_{\rm{data}}}-\ln{2\pi})
\end{equation}
Where $d$ is the number of free parameters of the fit, $N_{\rm{data}}$ is the number of data points and $\Delta \text{BIC} = \text{BIC}(1)-\text{BIC}(2)$ for models 1 and 2. As can be seen, the BIC strongly penalises models with additional parameters. The second Bayes factor estimator is the Laplacian approximation (BF) given by:
\begin{equation}
\text{BF} = \frac{P(D|M1)}{P(D|M2)} = \frac{\int{P(D|P1, M1)\times P(P1|M1)dP1}}{\int{P(D|P2, M2)\times P(P2|M2)dP2}}
\end{equation}
$P(D|M1)$ is the marginal likelihood for model 1, $D$ is the data, and $P1$ is the parameter of model 1. Under the assumption that the probability distribution of $P(D|P1, M1)$ has a well defined peak around the best-fitting parameters and the shapes of the distribution are approximately Gaussian, Laplace's approximation can be used \citep{b59} to simplify the integrals to:
\begin{multline}
\int{P(D|P1, M1) \times P(P1|M1) dP1} = P(D|M1)\\ = 2\pi^{d1/2}\times \sqrt{\Sigma 1} \times \exp({-\chi(1)^{2}/2)} \times P(P1|M1)
\end{multline}
$d1$\ is the number of free parameters in the model $M1$, $\Sigma 1$ is the covariance matrix for the best fitting model as determined by `mpfitfun', and the prior $P(P1|M1)$\ is a uniform prior defined by the limits on parameters $P1(1), P1(2)...P1(d1)$.

Models that have larger $P(D|M)$ values are preferred. For this method, the addition of extra model parameters is strongly penalised by the $P(P|M)$ term which, for uniform priors, drops significantly due to the additional volume probed by the parameter space in the more complicated model. The Bayes Factor and $\Delta\text{BIC}$ have the proportionality $-2\ \text{log}(\text{BF}) \approx \Delta \text{BIC}$ so the same criteria can be used to evaluate the strength of evidence of M1 vs M2 and vice-versa. The criteria are taken from \citet{b59} and are cast in terms of comparing model M1 to M2 in Table \ref{bayestab}.
\begin{table}
\caption{Using $-2\ \text{log}(\text{BF})$ and $\Delta \text{BIC}$ to compare evidence in favour of model one over model 2.}
\begin{tabular}{lcc}
\hline 
$-2\ \text{log}(\text{BF})$ and $\Delta \text{BIC}$ & Evidence favouring $M1$ over $M2$\\
\hline
0 to 2 & Not stong
\\2 to 6 & Positive
\\6 to 10 & Strong
\\$>10$ & Very strong/decisive\\
\hline
\end{tabular}
\label{bayestab}
\end{table}

Aside from the two methods above, we incorporate the F-test using the algorithm of \citet{b60}. This tests whether the change in the $\chi^{2}$ value is significant given the change in the DOF for the more complicated model and returns a P-value as an indicator of significance. For the purposes of model selection, only when  $p<0.01$ is $M1$ favoured over $M2$. The three methods are used to give a single score (mod\_score) that can be used as an indicator that the more complicated model is favoured over the simpler model (e.g., a double-Gaussian over a single-Gaussian fit for H$\alpha$). The score is determined as, for $x = -2\ \text{log}(\text{BF}12), \Delta \text{BIC}$:
\begin{itemize}
\item If $2\leq x < 6$, mod\_score = mod\_score + 1
\item If $6\leq x < 10$, mod\_score = mod\_score + 10
\item If $x\geq 10$, mod\_score = mod\_score + 100
\end{itemize}
In addition to this if the $p$-value from the F-test is less than 0.01, then 1 is added to the mod\_score. For example, if both $-2\ \text{log}(\text{BF}12)$ and $\Delta \text{BIC}$ were greater than 10 and the F-test gave a $p$-value of less than 0.01 then the resultant mod\_score would be the maximum possible 201 indicating that model 1 is strongly favoured over model 2. As each model is compared to every other model, the comparison of the most complex of these models where the comparison score is 201 is the preferred model.

\section{AGN and galaxy pairs in GAMA}
\label{s3}
\subsection{Spectroscopic classification of AGN}
\begin{figure*}
\includegraphics[scale=0.78]{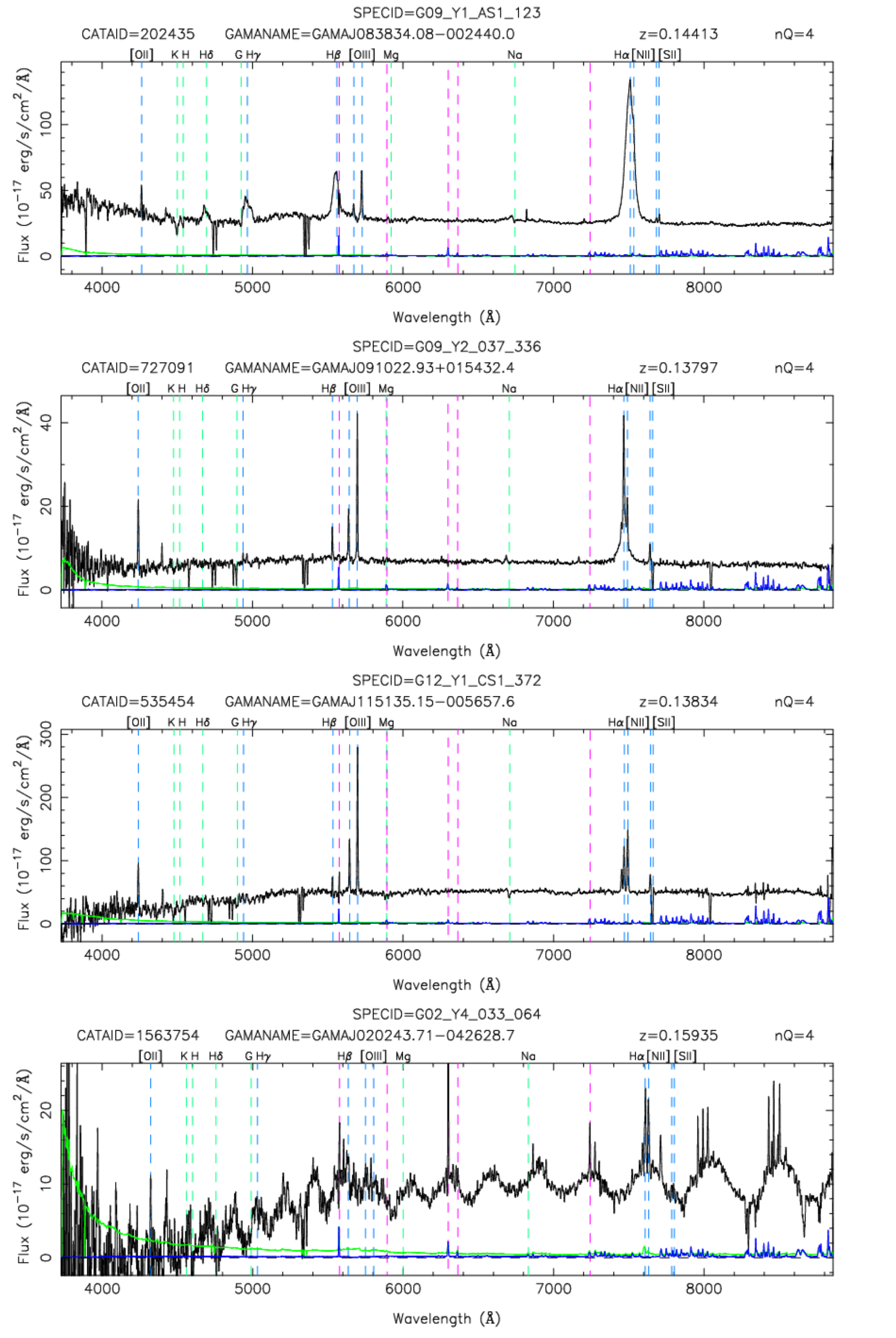}
\caption{From top to bottom: A typical `bona fide' T1 AGN spectrum; a typical `intermediate' type BLAGN spectrum; a typical T2 AGN spectrum; and example AAOmega spectrum affected by time dependent fringing, 19\% of our selected broad line spectra were discarded from our data as a result of this. These images are obtained from the GAMA \citep{b10, b46} `single object viewer' online tool.}
\label{spec}
\end{figure*}

The classification of AGN is not a discrete process, there exists a continuum between the truly broad line QSO's, which exhibit strong broad emission in H$\alpha$, H$\beta$, H$\gamma$ and H$\delta$, through to the narrow line T2s, which exhibit no observed broad component in their permitted emission lines (see Figure \ref{spec}). As such we create 3 catalogues of AGN based on the emission line properties taken from the GAMA emission line properties database (described in Section \ref{s2}). Only data from SDSS and AAOmega obtained spectra are used as only these spectra have been flux calibrated \citep{b33}. Catalogue 1 consists of `bona fide' type AGN that have broad emission components detected for both $\rm{H}\beta$ and $\rm{H}\alpha$. Catalogue 2 contains intermediate type AGN that show evidence for broad-line emission, but the total flux in the emission lines are dominated by the narrow component. These AGNs are often referred to as type 1.5, 1.8 and 1.9 AGN \citep{b29}. Our final AGN catalogue contains the T2 AGN that are classified based on their emission line ratios as per \citet{b18}. Example spectra from each of these catalogues is shown in Figure \ref{spec}. 

In order to ensure that the H$\alpha$ line is detected reliably by the AAOmega spectrograph we limit our study to a redshift of $z < 0.3$. To ensure only reliable spectra are used we select only those with $nQ \geq 3$ indicating that there is a $>90\%$ chance that the redshift is accurate \citep{b10, b46}. Furthermore, we require $\text{S/N}>3$, S/N is measured as $\text{Flux}/\Delta\text{Flux}$ for all emission lines required to classify a galaxy as an AGN type. To ensure in cases where the red and blue components of 2dF/AAOmega spectra have not been spliced cleanly that this splicing does not result in a false detection of a H$\beta$ broad line, we exclude all galaxies with $0.170<z<0.175$ from selection. In order to provide a clean detection of the H$\alpha$ line that is free from telluric contamination by the A-band Fraunhofer lines we also exclude galaxies with $0.157<z<0.163$. 

To select our bona fide T1 catalogue we select galaxies that have a broad component, i.e. a FWHM $\geq$ 1200 km s$^{-1}$, in H$\alpha$ and H$\beta$. This is selected for by requiring that for H$\alpha$ and H$\beta$ the complex double Gaussian model is preferred to more simple emission line fits. We require that the flux of the broad component of a line be greater than the flux from the narrow component. Also, in agreement with \citet{b29}, we select only galaxies where the H$\alpha$:H$\beta$ broad flux ratio is less than 5 to be classified as bona fide T1 AGN. For quality control purposes we select only from spectra where for both H$\alpha$ and H$\beta$ the amplitude and dispersion of the broad component are not pegged at the parameter boundaries.

To select our intermediate type AGN we require that at least the H$\alpha$ line has a broad component, i.e. that the complex double Gaussian model is preferred for this line. As with the T1 selection we require for quality control purposes that the amplitude and dispersion of the broad component Gaussian of H$\alpha$ to have not pegged at the parameter boundaries. Furthermore, to ensure that a broad H$\alpha$ line is the result of nuclear activity we require that there is a significant [O III](5007\AA) detection. The [O III](5007\AA) is considered to be an indicator of nuclear activity \citep{b37}. Specifically [O III](5007\AA) has been shown to be consistently present in AGN selected by hard X-ray emission \citep{b26} and indeed [O III](5007\AA) luminosity scales with X-ray luminosity suggesting it is a useful indicator of AGN power \citep{b26, b51}. We define a significant detection of \text{[O III](5007\AA)} in this context to be an equivalent width greater than 3\AA.

Approximately 5\% of AAOmega spectra suffer from a time-dependent fringing artefact \citep[shown in Figure \ref{spec}]{b33}. For the purposes of emission line modelling this can mimic the presence of a broad line. Therefore, to remove any affected spectra we have selected, the broad line AGN (BLAGN; T1bona fide and intermediate) catalogues were visually inspected for this and the affected spectra discarded. We found that 19\% of our broad line AGN were affected and therefore discarded. Further to this 6 of our intermediate type AGN were removed from consideration due to spectra that showed evidence of contamination by another object, e.g., a nearby bright star. 
\begin{figure*}
\includegraphics[scale=0.8]{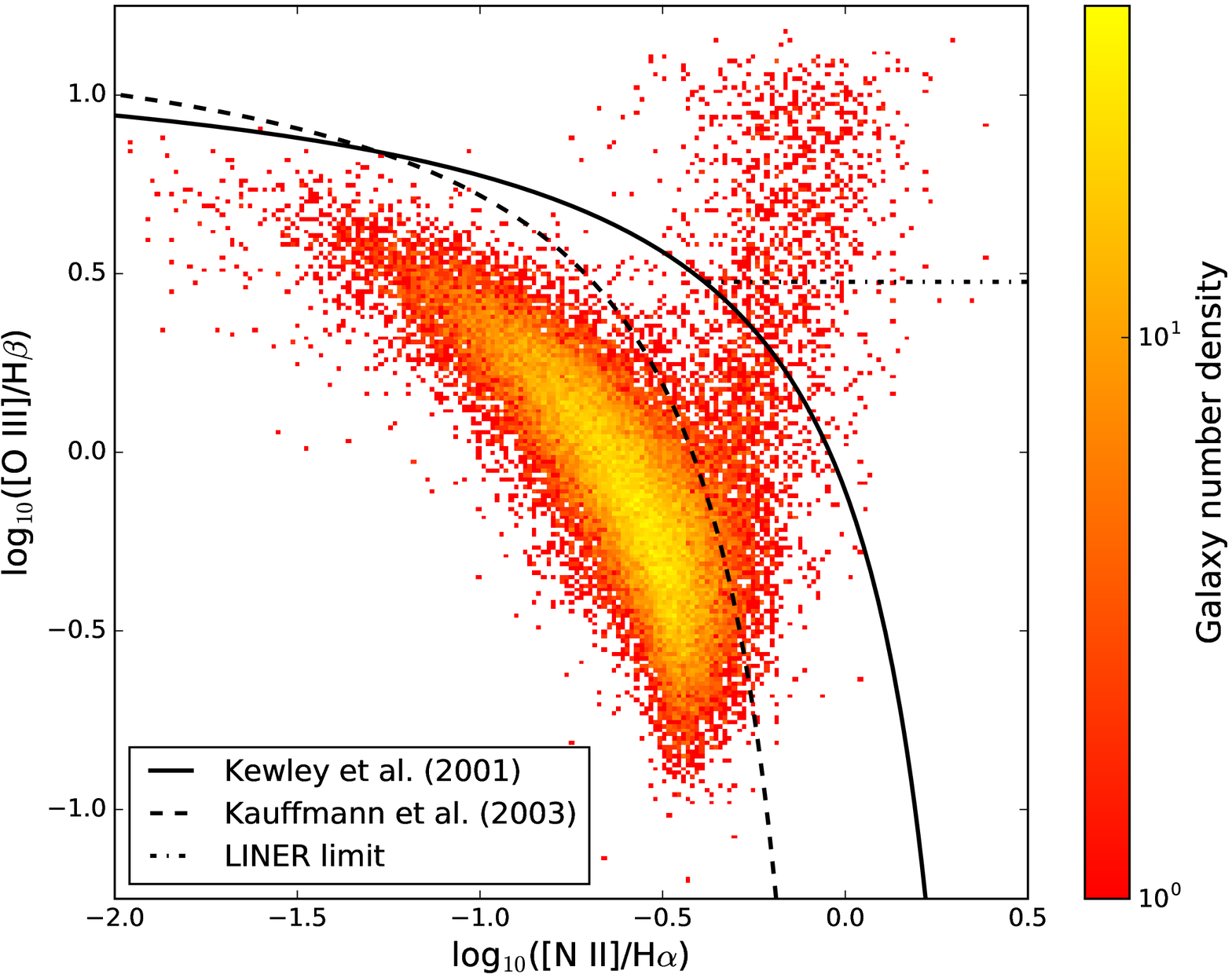}
\caption{2-dimensional histogram BPT plot of our galaxy sample. The black dot-dashed line is [OIII]/H$\beta=3$ to identify LINERs, the black solid curve is the \citet{b18} line to separate AGN from star forming galaxies and the black dashed line is the \citet{b19} line to segregate star-forming galaxies from AGN. The subtle difference between the objective of the \citet{b18} and \citet{b19} lines leaves a region between these two curves where the line ratios can be explained by a combination of star-forming and nuclear activity. For our data we classify galaxies above both the \citet{b18} and $\text{[O III]}/\text{H}\beta = 3$ lines as T2 AGN.}
\label{bpt}
\end{figure*}

To select the T2 catalogues we select from those galaxies not selected as BLAGN and require that the amplitude, dispersion and position of the narrow Gaussian have fitted successfully for H$\beta$, \text{[O III]}(5007\AA), H$\alpha$ and [N II](6583\AA). We correct the flux measurements of the Balmer emission lines for stellar absorption as per the method of \citet{b33}: 
\begin{equation}
\rm{F}_{\rm{cor}} = \Bigg(\frac{\rm{EW}+2.5\AA}{\rm{EW}}\Bigg)\rm{F}_{\rm{obs}}
\end{equation}
Where:
\\$\rm{F}_{\rm{cor}}$ is the corrected flux measurement,
\\$\rm{F}_{\rm{obs}}$ is the observed flux measurement,
\\$\rm{EW}$ is the measured equivalent width of the emission line.
\\This fixed 2.5\AA\ equivalent width correction is found to be appropriate after comparison of the Gaussian fits to the emission lines in GAMA data with the {\sc gandalf} v1.5 \citep{b21} fits of the spectra which intrinsically account for stellar absorption. A full description of this comparison is given in Section 6.4 and, in particular, shown in Figure 13 of \citet{b33}. Furthermore, to test if this might bias our AGN selection, we compare differences in the position on a Baldwin-Philips-Terlevich (BPT; \citealp{b17}) diagram for our measurements versus those from the MPA/JHU SDSS emission line catalogue \citep{b74}. For galaxies that lie in the AGN region of the BPT diagram, the median difference in the ratios $\text{log}_{10}(\text{[N II]}/\text{H}\alpha)$ and log$_{10}(\text{[O III]}/\text{H}\beta)$ between our corrected and the MPA/JHU SDSS catalogues are -0.01 and -0.02, with scatter 0.02 and 0.06, respectively. The scatter in these differences is comparable to the median of the standard errors on the distributions of those ratios (0.02 and 0.05 for log$_{10}(\text{[N II]}/\text{H}\alpha)$ and log$_{10}(\text{[O III]}/\text{H}\beta)$, respectively). Thus, we conclude that no substantial bias in our T2 selection is induced by using the fixed 2.5\AA\ correction for stellar absorption.

We measure the ratios of the [N II](6583\AA) to H$\alpha$ flux and \text{[O III](5007\AA)} to H$\beta$ flux in order plot a BPT diagram (see \text{Figure \ref{bpt}}). The ionising radiation emitted by an AGN is harder than that produced by star-formation and, therefore, the two ionising sources excite line species at different rates. These differences manifest themselves in the BPT diagram (Figure \ref{bpt}), which reveals that star-forming galaxies and AGN lie in relatively distinct regions. \citet{b18} defined AGN on this diagram as having:
\begin{equation}
\log \bigg(\frac{\text{[O III](5007\AA)}}{\rm{H}\beta}\bigg) > \frac{0.61}{\log \big(\frac{\text{[N II](6583\AA)}}{\rm{H}\alpha}\big)-0.47} +1.19
\label{Kewleyeq}
\end{equation}
where H$\beta$, [O III](5007\AA), H$\alpha$ and [N II](6583\AA) refer to the flux measurements of those emission lines.

The \citet{b18} criterion is a conservative segregator of nuclear from star-forming activity. Ergo, some of the galaxies that are close, but fail to satisfy this criterion may be star forming galaxies that also host an AGN. This criterion also selects galaxies with low ionisation nuclear emission regions (LINERs) which are often defined \citep{b19} as having:
\begin{equation}
\frac{\text{[O III](5007\AA)}}{\text{H}\beta} < 3
\label{linereq}
\end{equation}
\\Where H$\beta$ and [O III](5007\AA) refer to the flux of the hydrogen beta and [O III](5007\AA) emission lines respectively. Although sometimes considered to be a subclass of weak AGN, there is some controversy over the nature of LINERs. Recent evidence suggests that black hole accretion may not sufficiently explain these objects, with photoionisation by an ageing stellar population being invoked as a possible explanation \citep{b65}. Further to this,  spatially resolved integral field unit spectroscopy observes LINER emission to be extended across kpc scales within a galaxy and not just confined to the nucleus \citep{b20, b64}. This further suggests a non-nuclear origin for the ionisation within these galaxies. As such we exclude these galaxies from our sample of AGN, that is our sample of T2 AGN must satisfy equation \ref{Kewleyeq} but not equation \ref{linereq}.

Obtaining stellar mass estimates for AGN is non-trivial. The GAMA stellar mass catalogue \citep{b25} provides stellar masses for every galaxy within the survey as obtained through modelled photometry (for a full description see \citealp{b25}).
However, for some AGN, particularly QSOs and T1 AGN, the power-law contribution to the continuum may dominate over the stellar component (note the increase in continuum flux at the blue end of the T1 spectrum in the top panel in Figure \ref{spec}), making the stellar mass measurements unreliable.
The [O III](5007\AA) emission line is expected to occur due to nuclear ionisation of the narrow line region and acts as a proxy for AGN power \citep{b26, b37}. Thus for our analysis we decide to match our AGN on \text{[O III](5007\AA)} luminosity rather than stellar mass when comparing between AGN types. Furthermore, we note that it is very difficult to correct \text{[O III](5007\AA)} for extinction due to contamination of the observed Balmer decrement by the broad line region \citep{b19}. This issue is further compounded by the differing spatial scales likely to be associated with the broad Balmer and \text{[O III](5007\AA)} emission regions in the galaxy, hindering the accuracy of any dust correction calculations based on the Balmer decrement from the broad line region. Consequently we compare AGN of different types by their observed \text{[O III](5007\AA)} luminosity.

\begin{figure}
\includegraphics[scale=0.45]{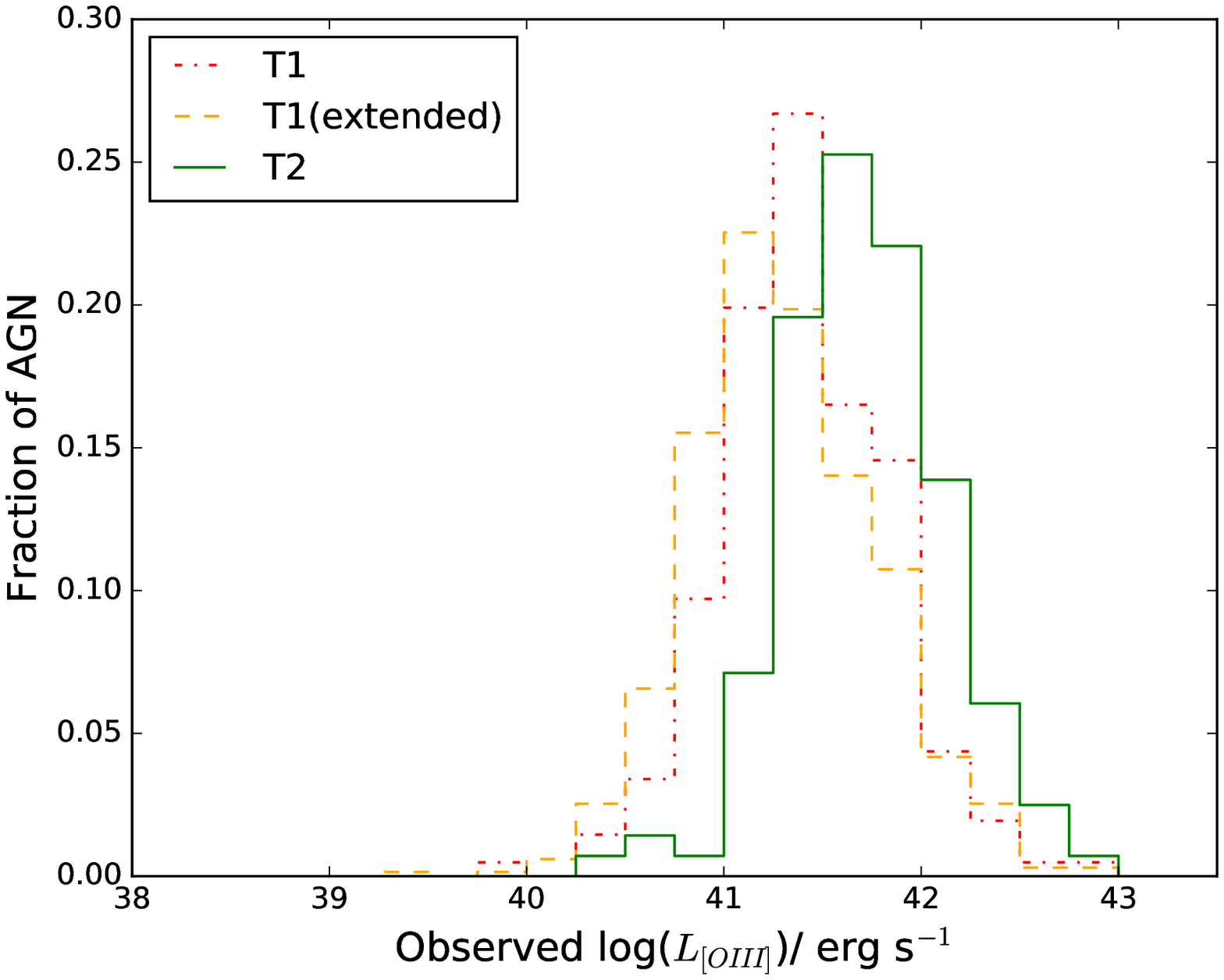}
\caption{The distribution of observed (not extinction corrected) \text{[O III](5007\AA)} luminosities in our sample by AGN type.}
\label{odist}
\end{figure}

For our work we choose to group our AGN into three \text{[O III](5007\AA)} luminosity bins: low ($\text{log}_{10}(L_{\text{[O\ III]}})\leq41.5\,\text{erg s}^{-1}$), mid ($41.5\,\text{erg s}^{-1}<\text{log}_{10}(L_{\text{[O\ III]}})\leq42\,\text{erg\ s}^{-1}$) and high ($\text{log}_{10}(L_{\text{[O\ III]}})>42\,\text{erg s}^{-1}$). These bins were to chosen to approximately split the populations into equal subsets while maintaining numbers in each separation bin when AGN in galaxy pairs are considered (see Section \ref{s41}). The observed \text{[O III](5007\AA)} luminosity distributions of the T1, extended T1 and T2 AGN samples are shown in Figure \ref{odist}. We note the higher \text{[O III](5007\AA)} luminosity distribution of our T2s compared to our T1s and extended T1 catalogue. We attribute this to the weaker dependence on the strength of \text{[O III](5007\AA)} emission line in the selection of these catalogues, and indeed total lack of dependence on this emission line in the case of the bona fide T1 catalogue.

Given the multiple spectra available for some of the galaxies within GAMA we use the spectrum with the most reliable redshift from a particular source. By proxy this subsets only the best quality spectrum for each galaxy as detected by a particular survey. That is to say if a particular galaxy had, for example, 5 spectra obtained within the catalogue, say 2 from GAMA, 2 from SDSS and 1 from 2dFGRS, we would be left with 3 spectra for this galaxy. The 2dFGRS spectrum and the best quality SDSS and GAMA spectra. Furthermore, we only use spectra from either SDSS or GAMA as within our catalogue only these are flux calibrated and hence only these will provide reliable measurements for the [O III](5007\AA) luminosity. We select our T1 catalogue first then remove the GAMA catalogue ID (CATAID) of that galaxy from consideration for the other two catalogues. That is to say if a spectrum is selected as a bona fide T1, then it cannot be reselected as an intermediate type or T2 AGN whose selection criteria it may also satisfy. We then remove any galaxies selected as intermediate type AGN from consideration as T2 AGN. To ensure each galaxy is only counted once in each catalogue, any duplicate CATAIDs (e.g. a GAMA and an SDSS spectrum for a particular galaxy) are removed from the catalogues.

This method is chosen over the more simple method of only selecting the spectrum with the most reliable redshift overall (regardless of survey) before applying our AGN selection criteria. We take this approach in order to ensure that some borderline AGN are not selected against due to a line of consequence being less well observed in the most reliably redshifted spectrum. When these two approaches are compared only 7 AGN are not selected by the simpler method that are selected by our method.

\begin{figure}
\includegraphics[scale=0.45]{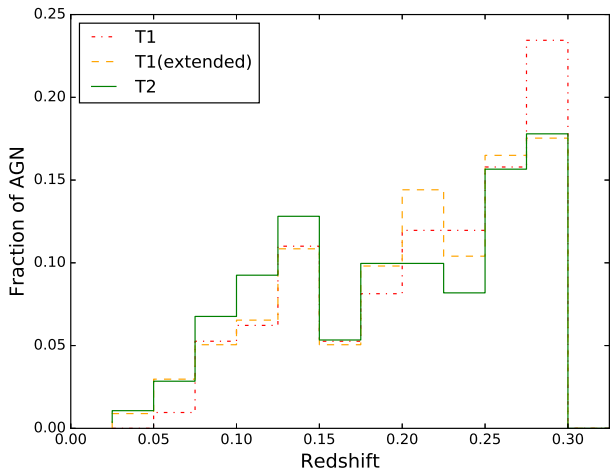}
\caption{The redshift distribution of the AGN in our sample by AGN type.}
\label{zdist}
\end{figure}

In total we find 954 AGN of all types across 4 catalogues: bona fide T1s; $1<\text{T}<2$; an extended catalogue which includes any AGN with a broad H$\alpha$ component, i.e. the $1<\text{T}<2$ catalogue appended to the bona fide T1 catalogue; T2. As a result of our selection of only AGN with high S/N on all emission lines required for classification, these numbers are lower than might be expected. Indeed, \citet{b69} have shown that using such unambiguous detections of all required emission lines may exclude up to half of the the AGN population that do not satisfy such strict selection criteria, consequently our AGN selection is conservative. The numbers of galaxies in each of our AGN catalogues is shown in Table \ref{agntab}. All of the AGN are selected from the three 60 square degree GAMA equatorial fields (G09, G12, G15). We note that our AGN types have similar redshift distributions, inseparable by use of a Kolmogorov-Smirnov (KS) test, enabling a fair comparison of the environment of these catalogues. The redshift distributions of our AGN are shown in Figure \ref{zdist}.
\begin{table}
\caption{The final number of selected AGN of each type in our sample. Also included is the number LINERs detected in our data selection.}
\begin{tabular}{lcc}
\hline 
AGN catalogue & No. of galaxies\\
\hline
T1 (bona fide) & 209
\\$1<\text{T}<2$ (intermediate type) & 464
\\T1 (extended; bona fide + intermediate catalogues) & 673
\\T2 & 281
\\LINER & 111\\
\hline
\end{tabular}
\label{agntab}
\end{table}

We further to this find 111 LINERs which satisfy the \citet{b18} criteria but have an [O III](5007\AA)/H$\beta$ flux ratio of less than 3. Our sample contains fewer LINERs than might be expected given that this class of galaxy may account for as much as one third of the local galaxy population \citep{b28}. We attribute this to two factors. Firstly, as these are weak emission line galaxies they are intrinsically harder to detect when a good signal to noise cut is applied. Indeed, in our sample of LINERs the median H$\beta$ equivalent width is less than 3\AA\ and thus only the highest quality spectra will detect these using our selection criteria. Secondly, the 2.5\AA\ Balmer absorption correction used is a general correction \citep{b33} and not optimal for the older stellar population associated with LINERs \citep{b65}.

\subsection{Selection of galaxy pairs}
In order to directly compare our results to those of \citet{b2}, we find all the GAMA galaxies within a projected separation $\text{d}R \leq$ $350\,\text{kpc }h^{-1}$ and redshift difference $|\Delta z| \leq 0.012$ of an AGN. This creates our pair catalogue, note that each AGN may be in a single pair, multiple pairs or not in a pair at all. The large limit placed on the redshift separation of our pairs prevents us from reliably selecting pairs from the GAMA group catalogue G$^{3}$Cv08 \citep{b12}.

These criteria result in a catalogue of 766 galaxies neighbouring 329 of our T1s (extended catalogue). Of these, 195 galaxies are neighbours to 81 of our bona fide T1s. For our T2 population we find that there are 273 neighbouring galaxies to 132 T2 AGN. The velocity separation used here is rather large. However, velocity difference is less significant than projected separation in finding true galaxy pairs \citep{b27}. This is shown in our sample, of which 79\% of the neighbours of the bona fide T1s, and 82\% of the neighbours of both the extended catalogue T1s and the T2s have $|\text{d}V| \leq 1000\,\text{km s}^{-1}$. When we take a subset of those neighbouring galaxies in `close pairs', i.e. within the GAMA group catalogue G$^{3}$Cv08 pair criteria \citep{b12, b45} of $\text{d}R\leq 100\,\text{kpc}\ h^{-1}$ and $|\text{d}V| \leq 1000\text{ km s}^{-1}$ we find that there are 28 neighbours to 22 of our bona fide T1s, 152 neighbours to 120 of the extended catalogue T1s and 59 neighbouring galaxies to 50 T2s.

\section{Analysis and Discussion}
\label{s4}
\subsection{T1 and T2 AGN fractions in pairs}
\label{s41}
\begin{figure}
\includegraphics[scale=0.55]{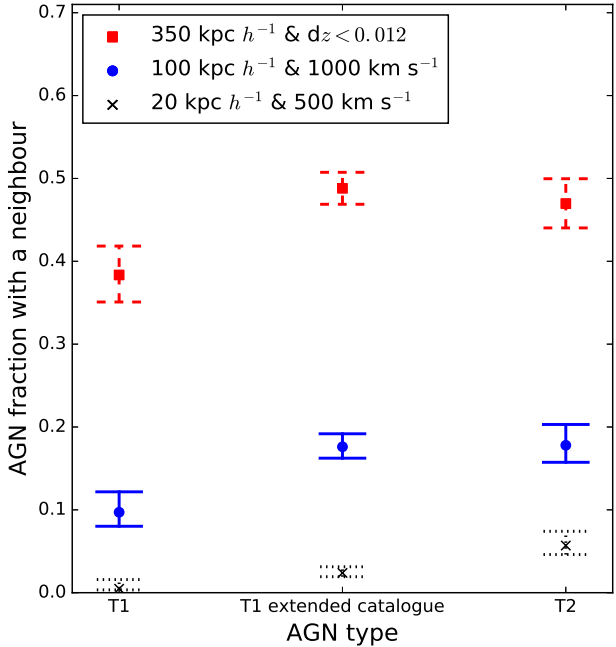}
\caption{The likelihood of AGN of each type to be in a pair or a close pair. The red squares use our own criteria of  $\text{d}R < 350\,\text{kpc}\ h^{-1}$ and $\Delta z < 0.012$. The blue circles use the GAMA definition of a pair \citep{b45} of $\text{d}R < 100\,\text{kpc}\ h^{-1}$ and $\text{d}R < 1000\,\text{km s}^{-1}$. The black crosses use only pairs which are likely to be directly interacting \citep{b45}, i.e., those with $\text{d}R < 20\,\text{kpc}\ h^{-1}$ and $\text{d}R < 500\,\text{km s}^{-1}$. We note that for the closest pair criterion (the black crosses) only 1 pair is found with a T1 and 16 pairs with T1(extended catalogue) and T2 AGN. This plot includes AGN of all \text{[O III](5007\AA)} luminosities and the error bars are binomial.}
\label{pairfrac}
\end{figure}

\begin{figure}
\includegraphics[scale=0.54]{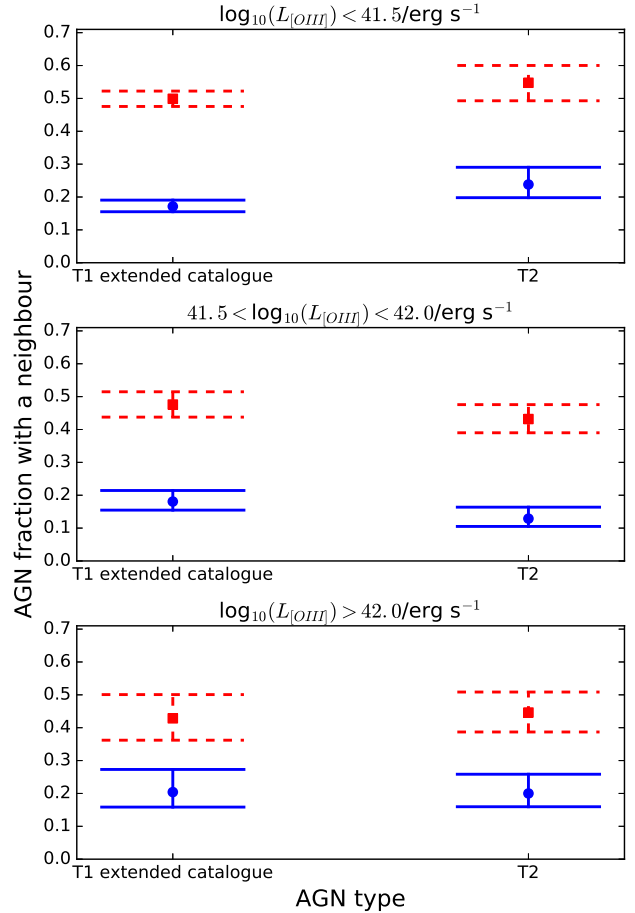}
\caption{The likelihood of AGN of each type to be in a pair or a close pair by luminosity bin. The legend is the same as in figure \ref{pairfrac} and due to low numbers the only the extended catalogue T1s compared to the T2s and pairs with $\text{d}R < 20\,\text{kpc}\ h^{-1}$ and $\text{d}R < 500\,\text{km s}^{-1}$ are not included.}
\label{pairfraclbin}
\end{figure}

Given the previous results calling in to question AGN unification \citep{b39, b41, b40, b44, b2, b35}, our ambition is to test these results as thoroughly as possible using the high spectroscopic completeness of GAMA. The pair fraction has been used heavily as a proxy for identifying the impact of environment due to galaxy-galaxy interactions in, e.g., triggering star-formation, AGN and for measuring the galaxy merger rate \citep{b61, b62, b63, b45}. This is the measure that frequently shows evidence of a difference between T1 and T2 AGN when such a result is found. Therefore, we compare the fraction of AGN with a neighbour within both the larger and tighter pair selection (see Figures \ref{pairfrac} and \ref{pairfraclbin}).

We find no significant difference in the fraction of T1 and T2 AGN found in a pair using our own pair criteria regardless of which T1 selection is used. This observation stands when only AGN in the same \text{[O III](5007\AA)} luminosity bin are compared. Furthermore, in contrast with the results of \citet{b39}, \citet{b40} and \citet{b35} we find no significant difference in the fraction of T1 and T2 AGN found in pairs with $\text{d}R<100\,\text{kpc }h^{-1}$ and $|\text{d}V|<1000\,\text{km s}^{-1}$. Again, this result does not change when only AGN with similar \text{[O III](5007\AA)} luminosity are compared. However, we do note that for both pair criteria there is suggestive, though still insignificant, evidence that the bona fide T1s may be marginally less likely to be found in a galaxy pair than AGN from either of the other two catalogues. With more data to reduce the statistical uncertainties, this may indeed support prior observations of more T2s with a close neighbour than T1s. The errors on this and the other fractional measurements throughout this paper are binomial assuming a beta distribution \citep{b38}.

Pair separations of $\text{d}R<100\,\text{kpc }h^{-1}$ and $|\text{d}V|<1000\,\text{km s}^{-1}$ or similar are commonly used in literature (e.g., \citealp{b45, b70}; the latter using $\text{d}R<80\,\text{kpc }h^{-1}$ and $|\text{d}V|<500\,\text{km s}^{-1}$) as this includes the scale of the Milky Way - Magellanic cloud system, and indeed these galaxies have most likely recently interacted or are currently interacting. In order to assess the fraction of AGN that are that are most likely to have undergone a recent interaction, or are currently interacting, would require an investigation of pairs separated by $\text{d}R<20\,\text{kpc }h^{-1}$ and \text{$|\text{d}V|<500\,\text{km s}^{-1}$} \citep{b45}. Given the relatively small size of our sample we are limited on this front, however we observe that $2.39^{+0.74}_{-0.46}\%$\ and $5.69^{+1.72}_{-1.09}\%$ of our extended catalogue T1s and T2s respectively are in these very tight pairs and are hence likely to be undergoing a merger. \citet{b67} found that less than 2\% of galaxies with $8.5 < \text{log}(\text{M}_{\star}/\text{M}_{\odot}) < 11$ are undergoing a merger at any one time, thus our observations suggest that there may be an excess of T2s undergoing mergers.
\begin{figure*}
\includegraphics[scale=0.39]{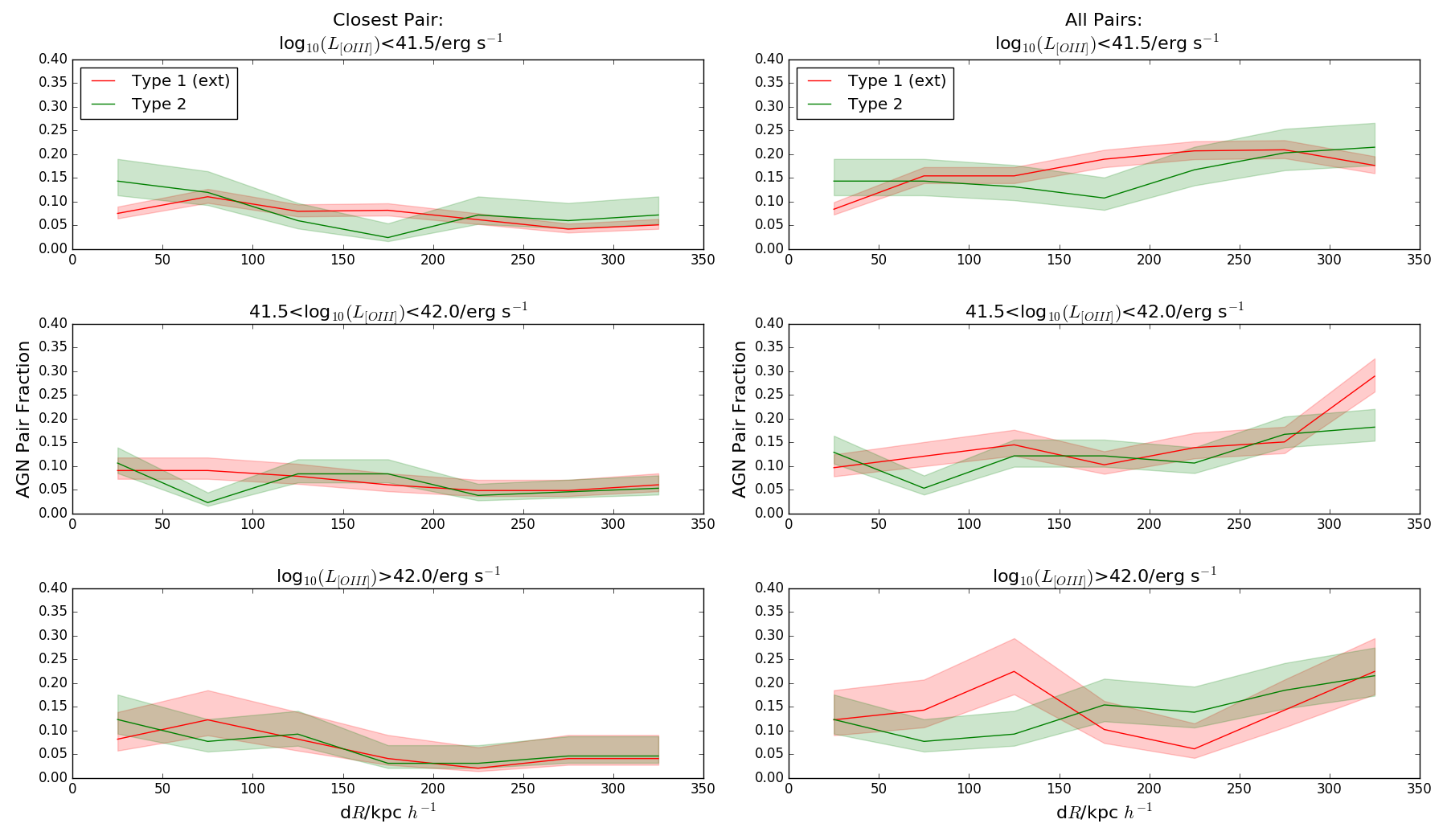} 
\caption{The fraction of our AGN sample found in pairs of galaxies by projected separation. Red lines represent the T1 (extended catalogue) AGN and green lines the T2s. The shaded regions are the one sigma binomial error limits. Left column: T1 (extended catalogue) and T2 AGN pair fractions by pair separation for only the closest pair (in projection) for each AGN, low to high $L_{\rm{[O\ III]}}$ AGN from top to bottom. Right Column: T1 (extended) catalogue and T2 pair fractions for all pairs satisfying our pair criteria, low to high $L_{\rm{[O\ III]}}$ from top to bottom.}
\label{pairfracdr}
\end{figure*}

We also calculate the pair fraction of our AGN as a function of pair separation. We separate our pairs into 50 kpc $h^{-1}$ bins and calculate the pair fraction in each bin. We define the pair fraction, $f$, as:
\begin{equation}
f = \frac{N_{\rm{pair}}}{N_{\rm{total}}}
\end{equation}
Where $N_{\rm{pair}}$\ is the number of AGN of that type and [O III](5007\AA) luminosity bin in a pair of the appropriate separation and $N_{\rm{total}}$ is the total number of AGN of that type and [O III](5007\AA) luminosity bin.
We compare the pair fractions of both the bona fide and extended T1 AGN catalogues with the T2 catalogue (see Figure \ref{pairfracdr}) and find no difference in the pair fractions with AGN type. We repeat this using only the nearest neighbour to each AGN such that each AGN is only used in one pair, the closest possible. Here we find no difference between either of our T1 populations and the T2s. As using either the T1 extended catalogue or the bona fide catalogue makes no difference to our results, we show the extended catalogue comparison with the T2s in Figure \ref{pairfracdr} to make use of higher AGN numbers and reduce the uncertainties.

\subsection{Neighbouring galaxies of AGN in pairs}
In order to test the colour differences of neighbouring galaxies to AGN found by \citet{b2}, we look at the properties of the neighbouring galaxies within our pairs. We use only those pairs where the neighbouring galaxy is not within any of our AGN or LINER catalogues, that is to say we exclude AGN-AGN and AGN-LINER pairs. We find that the vast majority of our AGN pairs are retained, with only 3, 9 and 4 AGN-AGN or AGN-LINER pairs respectively in our bona fide T1, extended T1 (including the bona fide T1s) and T2 AGN pair catalogues. With such low numbers, we are unable to investigate whether or not there are any trends in AGN-AGN pairs and restrict ourselves to AGN-non AGN pairs.

\subsubsection{Colours and stellar masses}
\begin{figure}
\includegraphics[scale=0.41]{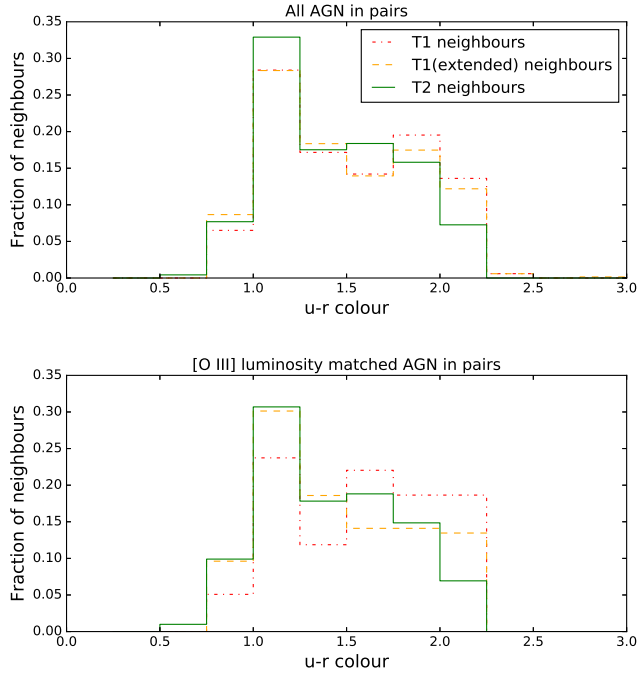}
\caption{Histograms showing the $u-r$ colour distributions of non-AGN neighbouring galaxies of AGN in pairs. Upper panel: all our AGN in pairs, lower panel: only AGN with $41.5< \text{log}_{10}(L_{\text{[O III]}}) \leq 42.0$ in pairs.}
\label{ur}
\end{figure}

We compare the $u-r$ colours of the neighbouring inactive (in a nuclear sense) galaxies of T1, extended T1 and T2 AGN in pairs. The $u-r$ colours are taken from the GAMA stellar mass catalogue ({\sc StellarMassesv18}; \citealp{b25}) and are taken from the modelled AB rest-frame SDSS $u$\ and $r$\ bands. These magnitudes are extinction and k-corrected. The distributions of the $u-r$ when the neighbours of AGN of all \text{[O III](5007\AA)} luminosities are considered appear to similar between AGN type (see Figure \ref{ur}). To statistically assess this apparent similarity we perform a KS test on the colour distributions of the neighbours of T2 and extended catalogue T1s to maximise the numbers of galaxies used in this analysis. We find that the probability that the colour distribution of the neighbouring galaxies of T1-extended and T2 AGN are drawn from the same parent population to be greater than 11\%, with a $p$-value of $0.117$. That is to say we cannot confidently say the distributions are drawn from different parent populations. This result doesn't change if the subset of neighbours of bona fide T1s are used instead of the neighbours of the whole extended T1 catalogue. When only AGN in our middle (and largest for numbers of AGN) [O III](5007\AA) luminosity bin ($41.5< \text{log}_{10}(L_{\text{[O III]}}) \leq 42.0$) are used, the KS derived $p$-value increases to 0.569, again indicating that the neighbouring galaxies of the AGN are likely to be drawn from the same parent sample. We apply this same test to the rest-frame, extinction-corrected $g-i$ colours of the neighbouring galaxies of AGN we find similar results. The results of the KS tests performed on the colour distributions are shown in Table \ref{colks}.

\begin{figure}
\includegraphics[width=8.5cm, height=9.3cm]{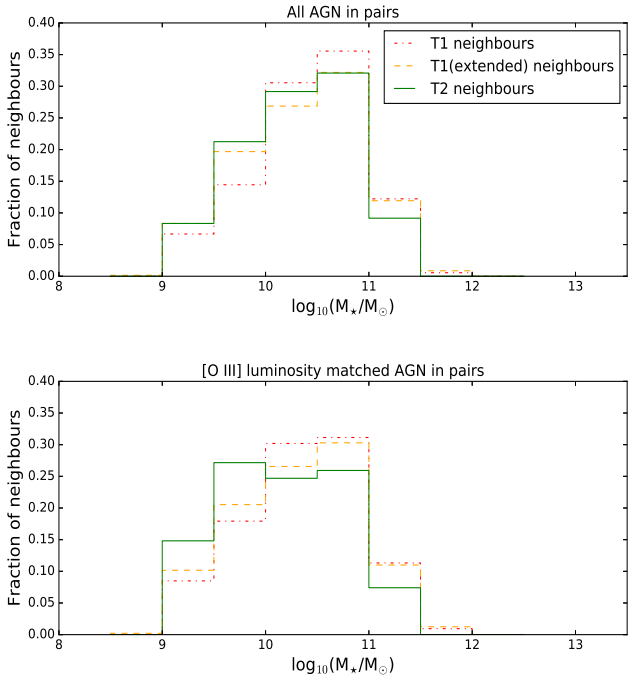}
\caption{Histograms showing the stellar mass distributions of the neighbouring galaxies to the AGN in pairs by AGN type. Upper panel: all our AGN in pairs, lower panel: only AGN with $41.5< \text{log}_{10}(L_{\text{[O III]}}) \leq 42.0$ in pairs.}
\label{massdist}
\end{figure}

We also compare the stellar masses of the neighbouring galaxies. The distributions are shown in Figure \ref{massdist}. These distributions are not separated by KS testing, giving a $p$-value of 0.324 when the extended T1 AGN neighbour masses are compared the T2 neighbour masses for all AGN \text{[O III](5007\AA)} luminosities.

\subsubsection{Star formation}
\begin{figure}
\includegraphics[scale=0.41]{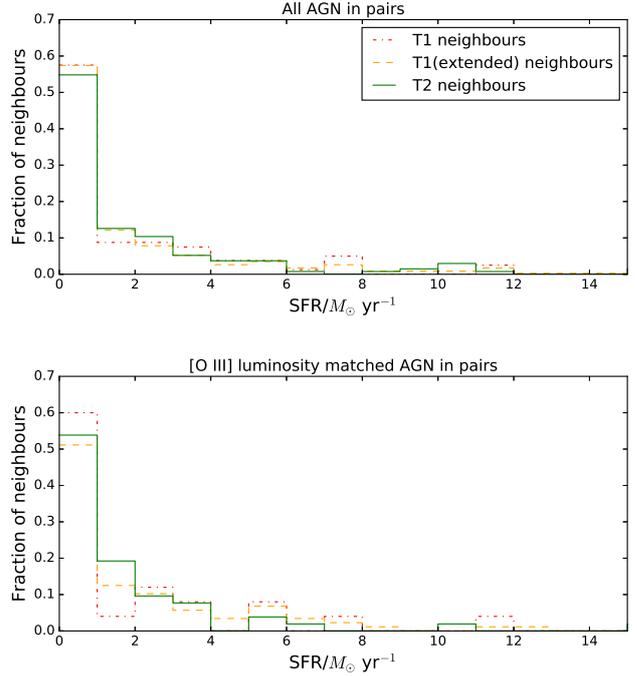}
\caption{Distribution of star formation rates for neighbouring galaxies of AGN in pairs that are not themselves AGN hosts and for which reliable emission line data is available. Upper panel: all our AGN in pairs, lower panel: only AGN with $41.5< \text{log}_{10}(L_{\text{[O III]}}) \leq 42.0$ in pairs.}
\label{sfr}
\end{figure}

\begin{figure}
\includegraphics[scale=0.41]{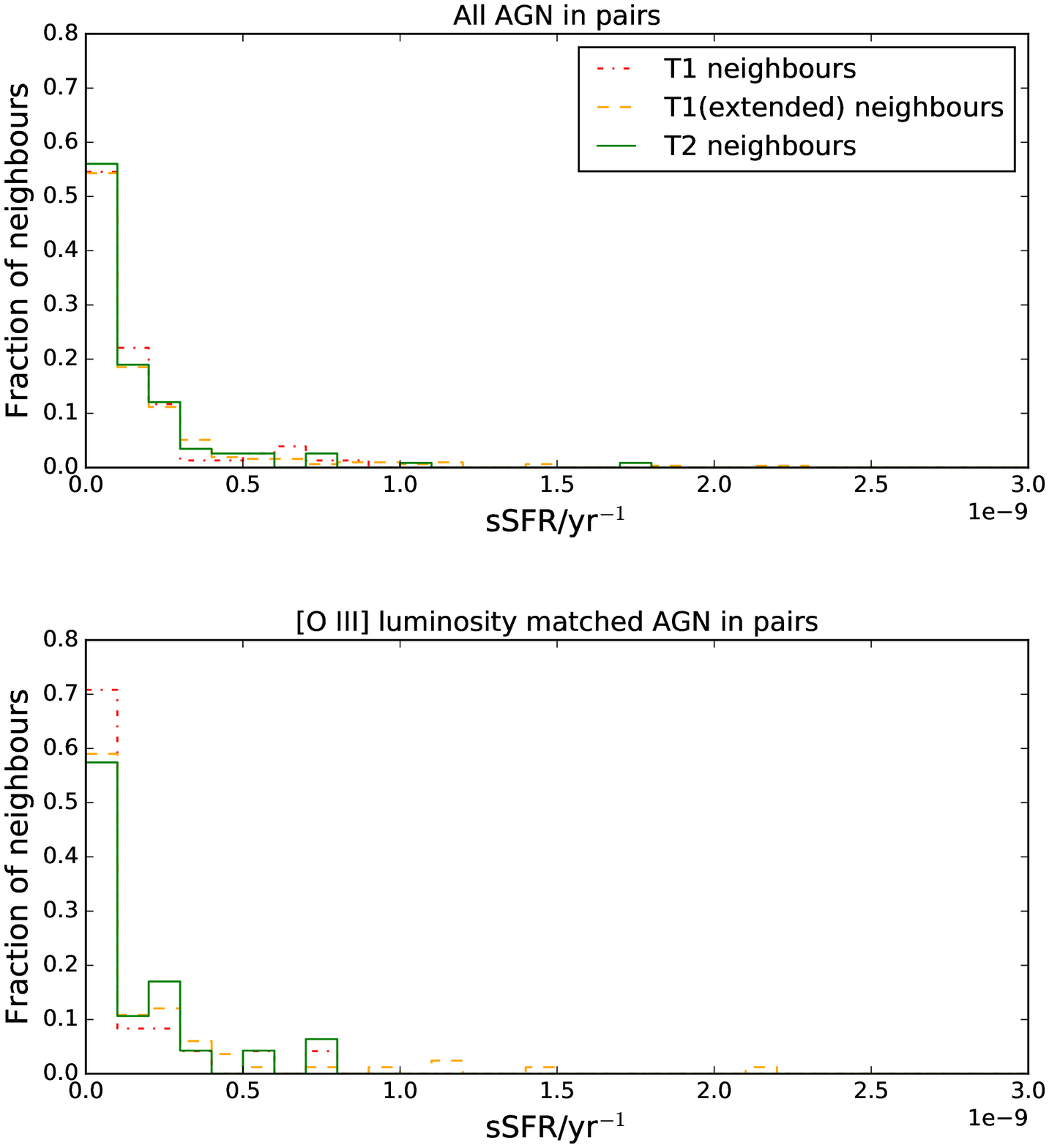} 
\caption{Distribution of specific star formation rates for neighbouring galaxies of AGN in pairs which are not themselves AGN hosts and for which reliable emission line data is available. Upper panel: all our AGN in pairs, lower panel: only AGN with $41.5< \text{log}_{10}(L_{\text{[O III]}}) \leq 42.0$ in pairs.}
\label{ssfr}
\end{figure}

Given the availability of spectra for all galaxies in our pairs, we compare the star-formation rates (hereafter SFR) of the neighbours where possible. This data is taken from the GAMA II emission line physical properties catalogue (to be produced as {\sc EmLinesPhysv05}; \citealp{b33}), where the star formation rate (in $\text{M}_{\odot}\ \text{yr}^{-1}$) assumes a \citet{b32} IMF and is found using:
\begin{equation}
\rm{SFR} = \frac{\textit{L}_{\rm{H}\alpha , \rm{int}}}{2.16 \times10^{34}}
\end{equation}
Where good mass data is available for the neighbouring galaxies this is taken from the GAMA stellar mass catalogue \citep{b25} and this is used to calculate the specific star formation rates (hereafter sSFR) where possible. The distributions of the SFRs and sSFRs are shown in Figures \ref{sfr} and \ref{ssfr} respectively. There is no apparent trend in SFR or sSFR of the neighbouring galaxy with AGN type and this is confirmed by performing KS tests on these distributions. The results of the KS tests for SFR and sSFR distributions are also shown in Table \ref{colks}. When only AGN of similar [O III](5007\AA) luminosity are used we still see no difference in either SFR or sSFR of the neighbouring galaxies.

\begin{figure}
\includegraphics[scale=0.45]{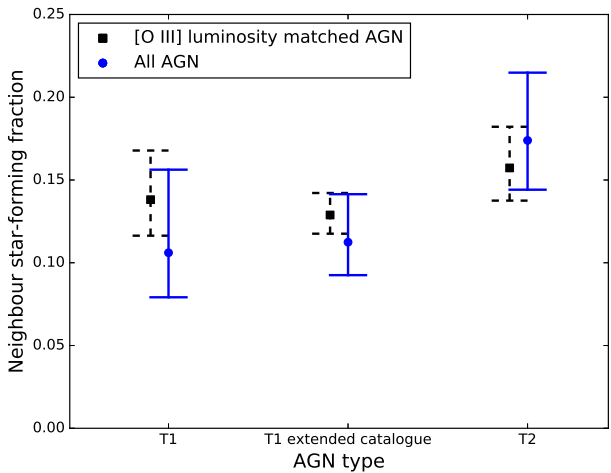}
\caption{Fraction of neighbouring galaxies of AGN that are classified as star forming by \citet{b19}. The blue circles with solid error bars represent all our AGN in pairs. The black squares with dashed error bars are representing AGN with $41.5< \text{log}_{10}(L_{\text{[O III]}}) \leq 42.0$ in pairs. The error bars are binomial and the two populations are horizontally offset from each other for clarity.}
\label{sf}
\end{figure}

We use the BPT criteria to classify the neighbours of the AGN, which are not themselves AGN or LINERs, as star-forming galaxies. We classify those neighbouring galaxies with all four emission lines with S/N > 3 and that satisfy the \citet{b19} criteria as star forming galaxies. We then compare the fraction of neighbouring galaxies classed as star formers by AGN type and find no difference in this fraction. This lack of difference in star-forming neighbour fraction is consistent regardless of [O III](5007\AA) luminosity bin matching. The fraction of star-forming neighbours by AGN type is shown in Figure \ref{sf}. These observations suggest that if the presence of an AGN is more likely in environments conducive to triggering star formation, then this does not differ with AGN type.

\subsubsection{5th nearest neighbour}
Finally we use the version 5 of GAMA environmental measures catalogue ({\sc EnvironmentMeasuresv05}; \citealp{b47}) to assess the environmental density of our AGN by type using the measurement to the fifth nearest neighbour ($\Sigma_5$). The environmental measures catalogue only uses galaxies with an absolute magnitude in the r band of $r_{\text{abs}}< -20$ and is limited to a redshift of $z<0.18$. This reduces the sample of AGN we can measure this for to 59, 219 and 112 for out T1 bona fide, T1 extended and T2 AGN catalogues respectively, of which 31, 147 and 77 have neighbouring galaxies. Upon using only AGN from our middle [O III](5007\AA) luminosity bin these numbers drop to 18 (6 with a neighbour) T1bona fide catalogue, 45 (27) T1 extended catalogue and 49 (33) T2.

Given the low numbers of bona fide T1s at $z<0.18$, we only compare AGN from the extended T1 catalogue with the T2s. The median $\Sigma_5$ values of the extended T1 and T2 AGN in the environmental measure catalogue are $1.92^{+1.23}_{-0.87}$ and $1.94^{+0.81}_{-0.89}\,\text{Mpc}$\ respectively, where the lower and upper bounds refer to the 25th and 75th percentiles. When we apply a KS test the $\Sigma_5$ distributions we find no difference between the extended T1 and T2 populations (see Table \ref{envks}). In order to compare the colours and (s)SFRs with the $\Sigma_5$ we redo these tests for only AGN with $z<0.18$. Here we again find that there is no significant difference in either the $u-r$, $g-i$, SFRs or sSFRs of the neighbouring galaxies of different AGN types.. A similar result is found when only AGN from our middle \text{[O III](5007\AA)} luminosity bin are compared. The full results of these KS tests are given in Table \ref{envks}. Our data suggests the T1 and T2 neighbours are drawn from the same population.
\begin{table}
\caption{Results of the 2 sample KS tests applied to the $u-r$, SFR and sSFR distributions of neighbouring galaxies of the extended T1 and T2 AGN. The $p$-values for all the AGN in pairs and those that a matched to others with $41.5< \text{log}_{10}(L_{\text{[O III]}}) \leq 42.0$ are shown.}
\begin{tabular}{lcc}
\hline 
Neighbour property & $p$(all $L_{\text{[O III]}}$) & $p$($41.5< \text{log}_{10}(L_{\text{[O III]}}) \leq 42.0$)\\
\hline
$u-r$ & $0.117$ & $0.569$
\\$g-i$ & $0.110$ & $0.736$
\\SFR & $0.676$ & $0.675$
\\sSFR & $0.833$ & $0.815$\\
\hline
\end{tabular}
\label{colks}
\end{table}
\begin{table}
\caption{Results of the 2 sample KS tests applied to the $u-r$, $g-i$, SFR and sSFR distributions of neighbouring galaxies of the extended T1 and T2 AGN with $z<0.18$ to enable comparison with the GAMA environmental measures catalogue. Also included are the KS test results of the $\Sigma_5$ measure for the AGN. The $p$-values for all the AGN in pairs and those that a matched to others with $41.5< \text{log}_{10}(L_{\text{[O III]}}) \leq 42.0$ are shown.}
\begin{tabular}{lcc}
\hline 
Neighbour property & $p$(all $L_{\text{[O III]}}$) & $p$($41.5< \text{log}_{10}(L_{\text{[O III]}}) \leq 42.0$)\\
\hline
$u-r$ & $0.0248$ & $0.199$
\\$g-i$ & $0.0119$ & $0.108$
\\SFR & $0.600$ & $0.914$
\\sSFR & $0.437$ & $0.758$
\\$\Sigma_5$ & $0.126$ & $0.508$\\
\hline
\end{tabular}
\label{envks}
\end{table}

\subsection{The effect of pair separation on neighbour properties}
\label{S4_pair_sep}
The use of any neighbouring galaxy within $\text{d}R<350\,\text{kpc}\ h^{-1}$ and $|\Delta z| \leq 0.012$ is somewhat liberal and is likely to include galaxies that are not truly interacting. To this end we take a subset of only those galaxy pairs that satisfy the more rigid criteria of $\text{d}R<100\,\text{kpc}\ h^{-1}$ and $|\text{d}V|<1000\,\text{km s}^{-1}$. This is chosen to approximate the scale of the Milky Way - Magellanic clouds interacting system \citep{b45}, and thus contain only interacting galaxies. Furthermore, we investigate the effect when only the closest neighbour of an AGN is used in this analysis rather than all neighbours satisfying the pair criteria. We redo all our neighbour property analysis for this close pairs data and find no differences in any of the neighbour properties when the extended T1 AGN are compared with the T2 AGN. The bona fide T1s are not used in comparison with the T2s here due to low numbers. We find no difference between T1 and T2 AGN when either of these pair criteria are used, using any of our neighbour property tests.

Of particular interest is the analysis when only pairs with $\text{d}R<100\,\text{kpc }h^{-1}$ and $|\text{d}V|<1000\,\text{km s}^{-1}$ are considered. Figure \ref{pairfracdr} in \citet{b45} shows that the fraction of galaxies that show signs of interaction in their morphologies increase as projected separation and, to a lesser extent, velocity difference decrease. Those galaxies that have $\text{d}R<20\,\text{kpc }h^{-1}$ and $|\text{d}V|< 500\,\text{km s}^{-1}$ are especially likely to show morphological disturbance. We show in Section \ref{s41} that there may be an excess of T2s in pairs this close relative to the extended T1s and indeed the general galaxy population \citep{b67}. We only have one bona fide T1 in a pair this close and as such do not compare the fraction of bona fide T1s to our other AGN populations here. \citet{b41} and \citet{b40} have previously suggested a model whereby galaxy interactions trigger star-formation before narrow line nuclear activity is produced during the interaction, with broad line nuclear activity occurring later, post interaction. Further still, the observations of \citet{b72} show that closer projected separation is linked to an increase in obscured AGN fraction, while \citet{b71} and \citet{b73} show an increased merger fraction in obscured AGN.

If AGN are preferentially obscured by close interactions and mergers (see in particular Figure 10 of \citealp{b71} and Section 5.2 of \citealp{b75}), then this could potentially be the result of a couple of different mechanisms. First, it may be the case that the either the gravitational effects of close interactions disturb the galaxy so as to disturb the morphology of the dust content of the torus and hence increase the chance of obscuration of an active nucleus. Second, the increased accretion onto the black hole could draw more dust toward the central engine and hence increase both the covering factor and opacity of the obscurer. If our observations hold up with a larger observed population they may provide supporting evidence for such models.

\section{Comparison with contrasting observations}
\label{S_comp}
Our observations are consistent with the the assumption that, being physically the same, T1 and T2 AGN reside in similar environments. On the smallest observed scales, our observations are consistent with the observations of \citet{b39, b40, b35} in that they show a possible excess of T2s in very close pairs. Such observations are viable within AGN unification should mechanisms such as those described in Section \ref{S4_pair_sep} be invoked.

The most puzzling aspect of our results is our inability to replicate the observations of \citet{b2} with our data. In particular, we find no difference in the colours of the neighbouring galaxies. As such this inevitably raises questions as to whether or not our comparison with \citet{b2} is a fair one, if ourselves and \citet{b2} are segregating our AGN in a similar manner, or if some other kind of selection effect may be biasing one of our results.

\subsection{The like colours of the neighbours of type 1 and 2 AGN}
The strongest colour difference in the neighbours of T1 and T2 AGN observed by \citet{b2} occurs when the pair is separated by less than $100\,\text{kpc}\,h^{-1}$. As our comparison of the colour distributions takes into account pair separations up to $350\,\text{kpc}\,h^{-1}$ it is logical to ask if our results change if we limit the analysis to sub $100\,\text{kpc}\,h^{-1}$ pairs. For this separation we only have 20 neighbours with reliable observed colours to out bona fide T1 AGN. As such, we only compare our extended T1 to our T2s in this test. Given that \citet{b2} select their T1 AGN only on the presence of a broad H$\alpha$ line, we consider this to be a fair comparison of our observations with those of \citet{b2}.

For AGN-galaxy pairs with $\text{d}R<100\,\text{kpc}\ h^{-1}$ and \text{$|\text{d}V|<1000\,\text{km s}^{-1}$} we find mean $u-r$ colours of the neighbours of the extended T1 and T2 AGN to be $1.49\pm 0.04$ and $1.49\pm 0.05$ mag respectively. If we limit our selected AGN to only those with \text{$41.5\,\text{erg}\,\text{s}^{-1}<L_{\text{[O III]}}<42.0\,\text{erg}\,\text{s}^{-1}$}, the mean neighbour \text{$u-r$} colours are $1.53\pm 0.07$ for the extended T1s and $1.56\pm 0.09$ for the T2s. The errors are estimated using the standard error of the mean. Given the scale of our errors, compared to the scale of the difference in colour observed by \citet{b2} ($\approx 0.3$ in their spectroscopic sample, rising to $\approx 0.5$ in their photometric sample), then we would expect to see a similar colour difference in our sample should one exist. Furthermore, the KS derived $p$-values for the colour distributions of the neighbouring galaxies of the extended T1 and the T2 AGN being statistically similar to be 0.989 and 0.910 for all AGN of all \text{[O III](5007\AA)} luminosities and $41.5\,\text{erg}\,\text{s}^{-1}<L_{\text{[O III]}}<42.0\,\text{erg}\,\text{s}^{-1}$ respectively. That is to say, when we observe the pairs where \citet{b2} found the strongest difference in colour between the neighbours of T1 and T2 AGN, we find no such difference.

\subsection{AGN selection}
\citet{b2} select their T1 AGN based solely on having a broad H$\alpha$ line, defined in their paper as $\sigma>10\,\AA$, based on the fit of a single Gaussian to the emission line. In the case of a broad emission line this may inaccurately model the shape of the emission line and hence provide inaccurate measurements on properties such as equivalent width. Further to this, the selection of T1 AGN based solely on H$\alpha$ likely includes partially obscured AGN (similar to our extended T1 selection). As we have shown, at least for our data, the partially obscured AGN population behaves similarly to the unobscured population (our bona fide T1 selection). As such we do not expect that this is the cause of the discrepancy between our observations.

There are also differences in how we select our T2 AGN. More specifically, like \citet{b2} we use \text{equation \ref{Kewleyeq}} \citep{b18} to select our T2s. This is a conservative measure of nuclear activity and excludes composite star-forming galaxies with AGN component as well as some LINERs. However, in order to ensure the fidelity of our T2 sample, we further to this exclude as a LINER any galaxy that satisfies equation \ref{linereq}.
\citet{b2} describe their method of selecting T2s in section 1.1.3 of the supplementary information of their paper, and how they remove LINERs in section 1.1.1. As they use the \citet{b19} criteria to select their T2s. Consequently, their T2 population may be more complete than ours, but may suffer from contamination by star-forming and composite SF-AGN galaxies. Further to this our exclusion of LINERs appears to be done using a different methodology. As such, it may the case that \citet{b2} and ourselves have T2 samples that are not drawn from the same parent population.

\section{Summary and Conclusions}
\label{s5}
In this work we have analysed the environment of 954 spectroscopically-selected AGN from the GAMA II galaxy survey \citep{b10, b46}. Exploiting the high spectroscopic completeness of GAMA \citep{b11} allows us to test recent observations suggesting different types of AGN reside in differing environments \citep{b39, b41, b40, b2, b35}. Thus, using GAMA we are able to comprehensively question the validity of the simple AGN unification scheme proposed by \citet{b4}.

Having compared the fraction of T1 and T2 AGN found in pairs we find no significant differences, contrasting with previous works \citep{b39, b40, b35}. This similarity in environment is maintained when our AGN-galaxy pairs are binned by projected separation and AGN [O III](5007\AA) luminosity. Taking the case of the simple AGN unification scheme, should orientation of the AGN relative to our line of sight be the only difference between T1 and 2 AGN, and should this orientation be approximately random, then one would expect to see no differences in the pair fraction of AGN by type. Furthermore, we observe no significant difference in the neighbours of AGN of different types. The $u-r$, $g-i$, stellar masses, SFRs and sSFRs of the neighbouring galaxies all have similar distributions, confirmed by KS testing, with AGN type.  Ergo, our observations here support the AGN unification model.

The one environmental difference we do note with AGN type occurs in galaxy pairs separated by less than $20\,\text{kpc}\ h^{-1}$ in projection. We find an excess of T2s in sub $20\,\text{kpc}\,h^{-1}$ pairs relative to the extended T1 and general galaxy population in such close pairs. One possible explanation for this could be that the simple AGN unification model holds except in the case of close gravitational interactions, where the geometry of the dust distribution in the galaxy is disrupted such that the likelihood of obscuration is increased. Alternatively, it may be that the interaction between galaxies drives more dust toward the nuclear region and it is this way that the probability of obscuration is increased.
In summary, our main findings are:
\begin{itemize}
\item The fraction of AGN found in pairs does not vary significantly with AGN type or pair separation down to $50\,\text{kpc}\ h^{-1}$. At separations of $< 20\,\text{kpc}\ h^{-1}$ and $500\,\text{km s}^{-1}$ there appears to be an excess of T2 AGN but more data will be required to confirm this.
\item The $u-r$ colour of the inactive neighbouring galaxies of AGN in pairs of galaxies does not appear to change depending on whether the AGN is of the broad or narrow line variety in contrast with \citet{b2}. The same result is found for the $g-i$ colours of the neighbouring galaxies to the AGN.
\item We find no difference in either the SFR or sSFR among the neighbouring galaxies of T1 and 2 AGN. We also find no difference in the fraction of these neighbours that are classed as star-forming according to a BPT diagram.
\item Comparing the distances to the 5th nearest neighbour of our AGN fails to find a difference between our broad and narrow line AGN populations.
\item Our results are generally consistent with the unified model of AGN proposed by \citet{b4} and further still support the observations of \citet{b72} and \citet{b71}.
\end{itemize}

The major hinderance to our observations has been our small sample size. GAMA II is highly complete allowing for spectroscopic observations of very close pairs of interacting galaxies. However, the survey footprint in the 3 equatorial regions (G09, G12, G15) is limited to $\approx 180$ square degrees. The small survey volume limits our ability to make statistically significant claims for well-matched AGN subsets, e.g., very close pairs with AGN of similar \text{[O III](5007\AA)} luminosity. Therefore, we state that although most of our results are consistent with AGN unification, the apparent excess in T2s in very close pairs with $\text{d}R<20\,\text{kpc }h^{-1}$ and $|\text{d}V|<500\,\text{km s}^{-1}$ suggest that AGN unification may not be a complete model. The question of AGN unification is thus likely to remain an open one for the foreseeable future.

To answer this will likely require a spectroscopically complete survey with a larger volume to provide a statistically robust complete spectroscopic analysis of AGN environment. Such a dataset may be available within the next few years from TAIPAN (Transforming Astronomical Imaging surveys through the Polychromatic Analysis of Nebula). A major component of TAIPAN will be the Taipan galaxy survey\footnote{www.taipan-survey.org} \citep{b34} which will use the TAIPAN spectrograph \citep{b22} on the UK Schmidt Telescope at the AAO. The Taipan galaxy survey aims to observe $\approx 1000000$ galaxies at $z<0.3$ with high spectroscopic completeness across the Southern sky. This survey is currently in the advanced planning phase and is expected to commence 4 years of operations in the first half of 2017. 

\section*{Acknowledgements}

The authors wish to thank the anonymous referee for their constructive comments. YAG acknowledges the financial support of the University of Hull through an internally funded PhD studentship that has enabled this research to be undertaken, as well as discussions with Jacob Crossett and Dane Kleiner. MSO acknowledges the funding support from the Australian Research Council through a Future Fellowship (FT140100255). MLPG acknowledges CONICYT-Chile grant FONDECYT 3160492.

GAMA is a joint European-Australasian project based around a spectroscopic campaign using the Anglo-Australian Telescope. The GAMA input catalogue is based on data taken from the Sloan Digital Sky Survey and the UKIRT Infrared Deep Sky Survey. Complementary imaging of the GAMA regions is being obtained by a number of independent survey programmes including GALEX MIS, VST KiDS, VISTA VIKING, WISE, Herschel-ATLAS, GMRT and ASKAP providing UV to radio coverage. GAMA is funded by the STFC (UK), the ARC (Australia), the AAO, and the participating institutions. The GAMA website is http://www.gama-survey.org/.

Funding for SDSS-III has been provided by the Alfred P. Sloan Foundation, the Participating Institutions, the National Science Foundation, and the U.S. Department of Energy Office of Science. The SDSS-III web site is http://www.sdss3.org/.

SDSS-III is managed by the Astrophysical Research Consortium for the Participating Institutions of the SDSS-III Collaboration including the University of Arizona, the Brazilian Participation Group, Brookhaven National Laboratory, Carnegie Mellon University, University of Florida, the French Participation Group, the German Participation Group, Harvard University, the Instituto de Astrofisica de Canarias, the Michigan State/Notre Dame/JINA Participation Group, Johns Hopkins University, Lawrence Berkeley National Laboratory, Max Planck Institute for Astrophysics, Max Planck Institute for Extraterrestrial Physics, New Mexico State University, New York University, Ohio State University, Pennsylvania State University, University of Portsmouth, Princeton University, the Spanish Participation Group, University of Tokyo, University of Utah, Vanderbilt University, University of Virginia, University of Washington, and Yale University.

This research made use of Astropy, a community-developed core Python package for Astronomy \citep{b36}.








%
%


\bsp	
\label{lastpage}
\end{document}